\documentclass[letterpaper]{article} 
\usepackage{aaai24}  
\usepackage{times}  
\usepackage{helvet}  
\usepackage{courier}  
\usepackage[hyphens]{url}  
\usepackage{graphicx} 
\usepackage{natbib}  
\usepackage{caption} 
\usepackage{algorithm}
\usepackage{algorithmic}

\usepackage{newfloat}
\usepackage{listings}
\DeclareCaptionStyle{ruled}{labelfont=normalfont,labelsep=colon,strut=off} 
\floatstyle{ruled}
\newfloat{listing}{tb}{lst}{}
\floatname{listing}{Listing}
\title{Non-Excludable Bilateral Trade Between Groups}
\author{
    Yixuan Even Xu,\textsuperscript{\rm 1}\thanks{This work was done when Xu was a visiting intern at CMU.}\quad 
    Hanrui Zhang,\textsuperscript{\rm 2}\quad 
    Vincent Conitzer\textsuperscript{\rm 3}
}
\affiliations{
    \textsuperscript{\rm 1}Tsinghua University,\quad  
    \textsuperscript{\rm 2}Simons Laufer Mathematical Sciences Institute,\quad  
    \textsuperscript{\rm 3}Carnegie Mellon University\\

    xuyx20@mails.tsinghua.edu.cn,\quad 
    \{hanruiz1,conitzer\}@cs.cmu.edu
}

\usepackage{enumerate}

\usepackage{amsthm}
\usepackage{amssymb}
\usepackage{amsmath, mathtools}

\usepackage{mleftright}

\usepackage{threeparttable}
\usepackage{hhline}
\usepackage{multirow}

\usepackage[capitalise]{cleveref}

\usepackage{tcolorbox}
\usepackage{caption}
\newfloat{algocf}{h}{lof}[section]
\crefname{algocf}{algorithm}{Algorithm}

\usepackage{tikz}
\usepackage{subcaption}
\usetikzlibrary{matrix}
\tikzset{mymatrix/.style={matrix of nodes, nodes={scale=0.8}}}

\usepackage{thmtools}
\usepackage{thm-restate}
\newtheorem{theorem}{Theorem}[section]

\newtheorem{corollary}{Corollary}[section]

\newtheorem{lemma}{Lemma}[section]

\newenvironment{proofof}[1]{{\bf Proof of #1:  }}{\hfill\rule{2mm}{2mm}}

\renewcommand{\d}{\mathrm{d}}
\newcommand{\R}{\mathbb{R}}
\newcommand{\U}{\mathcal{U}}
\newcommand{\N}{\mathcal{N}}
\newcommand{\vect}[1]{\ensuremath{\mathbf{#1}}}
\newcommand{\ind}[1]{\mathbf{1}\left[\vphantom{\sum}#1\right]}
\newcommand{\AutoAdjust}[3]{\mathchoice{ \left #1 #2  \right #3}{#1 #2 #3}{#1 #2 #3}{#1 #2 #3} }
\newcommand{\InParentheses}[1]{\AutoAdjust{(}{#1}{)}}
\newcommand{\InBrackets}[1]{\AutoAdjust{[}{#1}{]}}
\newcommand{\Ex}[2][]{\operatorname{\mathbf E}_{#1}\InBrackets{#2}}

\newcommand{\Prx}[2][]{\operatorname{\mathbf{Pr}}_{#1}\InBrackets{#2}}

\newcommand{\GFT}{\mathrm{GFT}}
\newcommand{\ALG}{\mathrm{ALG}}
\newcommand{\FB}{\mathrm{FB}}

\newcommand{\mechanism}{\mathcal{M}}

\newcommand{\bdist}{F}
\newcommand{\sdist}{G}

\newcommand{\dist}{F}

\newcommand{\disti}[1][i]{{\dist_{#1}}}

\newcommand{\val}{v}
\newcommand{\vals}{\vect{\val}}
\newcommand{\vali}[1][i]{{\val_{#1}}}

\newcommand{\bid}{b}
\newcommand{\bids}{\vect{\bid}}
\newcommand{\bidi}[1][i]{{\bid_{#1}}}
\newcommand{\bidsmi}[1][i]{\bids_{\text{-}#1}}

\newcommand{\cost}{c}
\newcommand{\costs}{\vect{\cost}}
\newcommand{\costj}[1][j]{{\cost_{#1}}}

\newcommand{\ask}{a}
\newcommand{\asks}{\vect{\ask}}
\newcommand{\askj}[1][j]{{\ask_{#1}}}
\newcommand{\asksmj}[1][j]{\asks_{\text{-}#1}}

\newcommand{\util}{u}

\newcommand{\utili}[1][i]{\util_{#1}}
\newcommand{\utilj}[1][j]{\util_{#1}}

\newcommand{\pay}{p}
\newcommand{\pays}{\vect{\pay}}
\newcommand{\payi}[1][i]{\pay_{#1}}

\newcommand{\rec}{r}

\newcommand{\alloc}{x}
\newcommand{\allocs}{\vect{\alloc}}
\newcommand{\alloci}[1][i]{\alloc_{#1}}

\begin{document}

\maketitle

\begin{abstract}
  Bilateral trade is one of the most natural and important forms of economic interaction: A seller has a single, indivisible item for sale, and a buyer is potentially interested.
  The two parties typically have different, privately known valuations for the item, and ideally, they would like to trade if the buyer values the item more than the seller.
  The celebrated impossibility result by Myerson and Satterthwaite
  shows that any mechanism for this setting must violate at least one important desideratum.
  In this paper, we investigate a richer paradigm of bilateral trade, with many self-interested buyers and sellers on both sides of a single trade who cannot be excluded from the trade.
  We show that this allows for more positive results.
  In fact, we establish a dichotomy in the possibility of trading efficiently. 
  If in expectation, the buyers value the item more, we can achieve efficiency in the limit. If this is not the case, then efficiency cannot be achieved in general.
  En route, we characterize trading mechanisms that encourage truth-telling, which may be of independent interest.
  We also evaluate our trading mechanisms experimentally, and the experiments align with our theoretical results.
\end{abstract}

\section{Introduction}
\label{sec:introduction}
Bilateral trade is one of the most natural and important forms of economic interaction in human society: A seller has a single, indivisible item
for sale, and a buyer is potentially interested.
The two parties typically have different valuations for the item, e.g., the seller has a production cost, and the buyer can derive a certain utility from consuming the item.
If the buyer values the item more than the seller, then trading in an appropriate way would benefit both parties: The buyer could offer to pay the seller a price strictly between both parties' valuations, giving both parties positive gain.
On the other hand, if the seller values the item more, then a trade would hurt the overall utility of them, and thus should not take place.
We can ensure that the trade takes place if and only if it is efficient by setting an appropriate price, {\em as long as both parties' valuations are public information}.

The situation is much subtler in the more realistic setting where the two parties' valuations are private.
There, maximizing the gain from trade becomes a {\em mechanism design} problem: We want to design a mechanism that, based on both parties' private valuations, decides whether they should trade, and if so, the amounts of money that the buyer pays and the seller receives.
Such a mechanism can be viewed as a trusted mediator that facilitates trade.
The key difficulty here is to encourage the buyer and the seller to truthfully reveal their private valuations, so that the mechanism can make the most socially efficient decisions, i.e., to trade whenever the buyer values the item more than the seller.

This basic and seemingly simple problem turns out to be surprisingly challenging.
In fact, the seminal impossibility result of \citet{myerson1983efficient} shows that in the above setting with private valuations, no mechanism simultaneously satisfies the following four natural properties:
\begin{enumerate}[\hspace{1em}$\bullet$]
    \item {\em (Bayesian) incentive compatibility}: In expectation, both parties maximize their respective utilities by truthfully reporting their valuations to the mechanism.
    \item {\em (Bayesian) individual rationality}: The expected values of participating are non-negative for both parties.
    \item {\em Weak budget balance}: If a trade happens, the amount of money that the buyer pays should be at least the amount that the seller receives, so that the mediator of the mechanism does not need to subsidize the trade.
    \item {\em Efficiency}: The two parties trade if and only if the buyer values the item more than the seller.
\end{enumerate}
Requiring incentive compatibility is WLOG given the revelation principle, so if we want to implement the socially efficient outcome, then we have to (1) force at least one party to reveal their true valuation, (2) force at least one party to participate, or (3) subsidize the trade --- none of which is feasible in practice.
This gives us one of the most famous impossibility results in economics: {\em Efficient bilateral trade cannot be implemented in a feasible way.}

Interestingly, the above impossibility result is only partially echoed by real-life phenomena.
Admittedly, negotiation between two individuals is often highly unpredictable and not fully efficient.
However, when we look at trade between two sizable ``organizations'' (e.g., countries) each consisting of many self-interested ``members'' (e.g., citizens), it appears that the aggregate decisions made by these members more often result in an outcome that overall benefits members of both organizations due to the large number of participants and is thus more efficient.
In such scenarios, the multiplicity of members in the two parties seems to provide a way around the classical impossibility result by \citet{myerson1983efficient}, opening up a new design space.

In this paper, we investigate this richer and seemingly more permitting paradigm of bilateral trade, with many self-interested members on both sides of the trade.
Roughly speaking, we consider a model where there are $n$ buyers and $n$ sellers on the two sides of the trade respectively\footnote{In most of our results, we treat $n$ as the only asymptotic variable. Therefore, we assume the two parties consist of the same number of agents $n$ for simplicity of presentation. Our results can also be adapted to the case when they are different.}, and all agents on each side share the same allocation and payment.
For concreteness, consider the following example: Amateur soccer club B wants to buy the right to use a certain soccer field from amateur soccer club S.
Members of the buying club B generally have different valuations for the right to the field, depending on how much each individual member enjoys training and playing.
These valuations are drawn independently from a common prior distribution.
The same also applies to the selling club S (with a different prior).
If the two clubs close a deal, all members of the buying club B get to enjoy the right to the field, and at the same time, they all make the same payment in the form of raised membership fees.
As for members of the selling club S, they lose the right to the field, but each member receives the same payment in the form of lowered membership fees.
In other words, members on each side of the trade share the same allocation and payment, and no one can be excluded from the trade.

In the above setting, ideally, the two clubs should trade if and only if on average, members of the buying club B value the right to the field more than members of the selling club S.
There are also other high-stakes real-world trading scenarios that are similar to the soccer club example, such as negotiations between countries, residential communities, or academic units (e.g., universities, schools, or departments).
In such settings, the preferences of all members should be accounted for, so naturally, a trading mechanism should take into consideration the reported valuations of each individual member of the two sides of the trade, and based on all this information, try to make the optimal decision for the society.
Our goal is to investigate the power and limitations of trading mechanisms in such settings, subject to participation and incentive constraints.

\subsection{Our Results}

We begin our investigation by looking at the gains-from-trade of the first-best mechanism. We show that when the buyers value the item more than the sellers in expectation, the first-best gains-from-trade (defined in \cref{sec:preliminaries}) is proportional to the number of agents, and when the sellers value the item (weakly) more, even the first-best gains-from-trade per agent goes to zero  (\cref{lemma:fb_values}). This naturally results in two different cases: when the buyers value the item more, trading in the right way can benefit each individual agent significantly, whereas when the sellers value the item more, there is not much value to be gained in the first place. We will approach these two cases separately.

\newpage
With this in mind, we proceed by first looking at deterministic trading mechanisms, which are both more practical and structurally simpler.
Our main result is a dichotomy: We show that if, in expectation, buyers value the item strictly more than sellers (i.e., we are in the first case discussed above), then essentially all desiderata discussed above can be achieved using a simple deterministic trading mechanism (Theorem~\ref{thm:forced_trade}).
Alternatively, if sellers value the item at least as much as buyers in expectation (i.e., we are in the second case discussed above), then feasible mechanisms, in general, can only achieve an approximation to the socially optimal outcome, by a factor strictly smaller than $1$. This holds even if we only require incentive compatibility (Theorem~\ref{thm:deterministic_hardness}).
So in short, the paradigm with multiple agents on both sides is in fact more permitting, but this makes a fundamental difference only when trading is beneficial in expectation (i.e., the first case). Nonetheless, when trading is not beneficial in expectation (i.e., the second case), there is little to be gained in the first place, so the fact that efficiency cannot be achieved in a feasible way does not cost much.

Technically, the above results are built on a powerful and intriguing characterization of feasible deterministic mechanisms: We show that any deterministic trading mechanism that that is incentive compatible (i.e., encourages truth-telling) must proceed in a voting-like way (Theorem~\ref{thm:deterministic_characterization}).
In particular, if we require strong budget balance, then there must be a unique, predetermined price independent of the agents' valuations (Theorem~\ref{thm:deterministic_characterization_sbb}).
The mechanism decides whether to trade by asking each individual agent whether they want to trade at the above price.
Moreover, in order to encourage truth-telling, this voting procedure has to be monotone (i.e., if the trade happens when a certain set of agents approve, then it must also happen when a superset of those agents approve).
More generally, this characterization might also be useful in other mechanism design problems that involve certain forms of non-excludability.

Given the dichotomy in the case of deterministic mechanisms, one natural question is whether randomization can help.
To this end, we characterize all incentive compatible randomized mechanisms that are ``well-behaved'' (i.e., twice continuously differentiable): Such mechanisms must be separable across different agents, in the sense that fixing the asks (resp.\ bids) of all sellers (resp.\ buyers), for each buyer (resp.\ seller), there is a component of the allocation function that depends only on the bid of that buyer (resp.\ seller), and the overall allocation function is the sum of all these components (Theorem~\ref{thm:randomized_characterization}).
Somewhat counterintuitively, this characterization implies an even stronger negative result: When sellers in expectation value the item more, no ``well-behaved'' randomized mechanisms that encourage truth-telling can achieve any constant approximation against the socially optimal outcome (Theorem~\ref{thm:randomized_hardness}).
In other words, randomization, at least when applied in a smooth way, cannot help in the paradigm of bilateral trade that we study.

Finally, we evaluate our trading mechanism with different classes of valuation distributions empirically.
The experimental results align with our theoretical results.
In particular, our mechanism consistently achieves all desiderata almost perfectly whenever the number of agents is large enough.

\section{Related Work}
\label{sec:related_work}

\paragraph{Bilateral trade.} Bilateral trade has been extensively studied in the classical setting where there is a single agent on each side of the trade. The seminal impossibility result of \cite{myerson1983efficient} shows that no mechanisms for bilateral trade can be both efficient and feasible (i.e., Bayesian incentive compatible, Bayesian individual rational, and weak budget balanced). As for approximation of gains-from-trade, \cite{mcafee2008gains} gives a $2$-approximation of the first best when the value distribution has a higher median than the cost distribution. \cite{blumrosen2016approximating} gives an $e$-approximation assuming a monotone hazard rate of the value distribution. \cite{brustle2017approximating} shows an unconditional $2$-approximation of the second best. \cite{deng2022approximately} shows an unconditional $8.23$-approximation of the first best, later improved into a $3.15$-approximation by \cite{fei2022improved}. Our results in this paper extend and complement existing research on bilateral trade.

\paragraph{Double auctions.} A related line of work to our setting is double auctions, which also generalizes bilateral trade to the case when there are multiple buyers and sellers. There is a significant body of work on double auctions \cite{babaioff2018best,babaioff2020bulow,doi:10.1137/1.9781611975482.11,brustle2017approximating,cai2021multi,mcafee1992dominant,colini2016approximately,colini2020approximately,colini2017fixed}. In the setting that we study, allocation and payment are shared across all agents, which is not the case in double auctions.

\paragraph{Non-excludable mechanism design.} Another related line of work is non-excludable mechanism design. Here, ``non-excludable'' means that if something is allocated, then it is allocated to all agents participating in the mechanism, i.e., no agent can be excluded from the allocation \cite{samuelson1954pure,mugrave1959theory,10.2307/1724745,ostrom1990governing}. A common example of non-excludable mechanisms is public projects, where the auctioneer allocates the same amount of the public goods to all players simultaneously. There is a rich body of work on this topic \cite{papadimitriou2008hardness,buchfuhrer2010computation,dobzinski2011impossibility,dughmi2016optimal,balseiro2021non}. In our setting, the allocation, payment, and receipt are similarly shared among all participants in each group. To the best of our knowledge, our work is the first to consider two sides trading in the context of non-excludable mechanism design with shared payment.

\section{Preliminaries}
\label{sec:preliminaries}

A set of $n$ sellers has a certain product that it can sell to a set of $n$ buyers. Each buyer $i\in[n]$ has a private value $\vali$, and each seller $j\in[n]$ has a private cost $\costj$, for if the trade happens. We denote the vectors of buyer values and seller costs as $\vals$ and $\costs$, respectively. According to the revelation principle, we can restrict our attention to incentive compatible mechanisms $\mechanism$, where each buyer $i$ submits a sealed bid $\bidi$ and each seller $j$ submits a sealed ask $\askj$. Let $\bids$ be the vector of all buyer bids and $\asks$ be the vector of all seller asks. The mechanism then decides an allocation $\alloc(\bids,\asks)\in[0,1]$ shared by all participants, indicating the probability of trading, a payment $\pay(\bids,\asks)\in\R$ shared by all buyers, and a receipt $\rec(\bids,\asks)\in\R$ shared by all sellers.

We study this multiplayer bilateral trade (MBT) problem in the Bayesian setting, where we assume that the buyer values $\vals$ are drawn i.i.d.\ from a known bounded prior distribution $\bdist$ and the seller costs $\costs$ are drawn i.i.d.\ from a known bounded prior distribution $\sdist$. Without loss of generality, let $F$ and $G$ be bounded on $[0,1]$. We also use $\bdist$ and $\sdist$ to denote the cumulative distribution functions, i.e., $F(\tau)=\Prx[\val \sim \bdist]{\val \leq \tau},G(\tau)=\Prx[\cost \sim \sdist]{\cost \leq \tau}\ (\tau\in[0,1])$.

For buyer $i$ and seller $j$, the quasilinear utilities are defined as $\utili(\bids,\asks) = \alloc(\bids,\asks)\cdot\vali -\pay(\bids,\asks)$ and $ \utilj(\bids,\asks) = \rec(\bids,\asks) - \alloc(\bids,\asks)\cdot\costj$ respectively. Incentive compatibility means that by truthfully revealing the private value or cost, each buyer or seller obtains the outcome maximizing his or her utility regardless of the bids and asks from others. For simplicity, we assume ties in utility are broken in favor of a trade. That is, if multiple outcomes share the same utility, the outcome with larger $\alloc$ is preferred by the participants. Then, in an incentive compatible mechanism, it is a dominant strategy to bid $\bids=\vals$ and ask $\asks=\costs$. As we consider an environment that requires shared allocation, payment, and receipt, we generalize the standard notion of individual rationality to one that considers each group as a whole, i.e., the total utility of buyers and sellers should be positive respectively.

For the bilateral trade problem, the mechanism designer is commonly concerned with the gains-from-trade (GFT), which refers to the expected utility gain from trading. Formally, for an incentive compatible mechanism $\mechanism$,
$
    \GFT = \Ex[\vals \sim \bdist^n, \costs \sim \sdist^n] {
        \InParentheses{\sum_{i} \vali - \sum_{j} \costj} \cdot \alloc(\vals,\costs)
    }.
$
To maximize GFT, ideally, we want the trade to happen whenever the total value of the buyers is greater than the total cost of the sellers. We call this optimal GFT the {\em first best (FB)}. That is, $\FB$ equals
\begin{equation*}
    \Ex[\vals \sim \bdist^n, \costs \sim \sdist^n] {
        \InParentheses{\sum_{i} \vali - \sum_{j} \costj} \cdot \ind{\sum_{i} \vali \geq \sum_{j} \costj}
    }.
\end{equation*}

We are interested in mechanisms that are incentive compatible (IC), individually rational (IR) in the above sense, weakly budget balanced (WBB), and \emph{efficient in the limit}, meaning that as $n\to\infty$, the mechanism's GFT $\ALG$ approaches $\FB$, i.e., $\frac{\ALG}{\FB}=1-o(1)$.

\section{Deterministic Mechanisms}
\label{sec:deterministic_mechanisms}

In this section, we consider deterministic mechanisms for multiplayer bilateral trade (MBT). Namely, the allocation function $\alloc(\bids,\asks)\in\{0,1\}$ is either $0$ or $1$. Before we proceed, we first present the following \cref{lemma:fb_values}. The proof follows standard calculation of concentration inequalities and is deferred to \cref{appsub:proof_of_lemma_fb_values}.

\begin{lemma}
    \label{lemma:fb_values}
    When $\Ex[\val\sim\bdist]{\val}\leq \Ex[\cost\sim\sdist]{\cost}$, $\FB = O(\sqrt{n})$. When $\Ex[\val\sim\bdist]{\val} > \Ex[\cost\sim\sdist]{\cost}$, $\FB = \Omega(n)$.
\end{lemma}

\cref{lemma:fb_values} naturally divides the MBT problem into two cases: when the buyers value the item more in expectation, and when the sellers value the item (weakly) more in expectation. In the latter case, the gains-from-trade per agent goes to zero even under the first-best mechanism.

With these two different cases in mind, we will approach them separately. We will first characterize the set of allocation functions that are implementable by an IC mechanism in \cref{subsec:deterministic_characterization}. Using this characterization, we establish a dichotomy in MBT: We will present a positive result when buyers value the item more in expectation (\cref{subsec:deterministic_positive}), and then show in the case that sellers value the item (weakly) more in expectation, no mechanism can, in general, achieve all the desiderata (\cref{subsec:deterministic_hardness}).

\subsection{Characterization of Incentive Compatible Mechanisms}
\label{subsec:deterministic_characterization}

The following \cref{thm:deterministic_characterization} is our characterization of the set of allocation functions that are implementable by an IC mechanism. If we additionally require the mechanism to be strongly budget balanced (SBB), i.e., the total payment of the buyers always equals the total receipt of the sellers, then it can be further simplified as described in \cref{thm:deterministic_characterization_sbb}.

At a high level, our characterization shows that any deterministic trading mechanism that is incentive compatible must proceed in a voting-like way.
In particular, if we require strong budget balance, then there must be a unique, predetermined price independent of the agents' valuations.
The way the mechanism decides whether to trade is by asking each individual agent whether they want to trade at the above price, which can also be viewed as running a voting procedure.
Moreover, in order to encourage truth-telling, this voting procedure has to be monotone (i.e., if the trade happens when a certain set of agents approve, then it must also happen when a superset of those agents approve).

\begin{theorem}
	\label{thm:deterministic_characterization}
	A deterministic allocation function $\alloc(\bids,\asks)$ for MBT can be implemented by an IC mechanism if and only if both of the following statements hold.
    \begin{enumerate}[\hspace{1em}\normalfont(a)]
        \item $\forall \asks \in [0,1]^{n}$, there exists $\tau_\asks \in [0,1]$ and a monotone Boolean function $f_\asks:\{0,1\}^{n}\to\{0,1\}$, such that $\alloc(\bids,\asks) = f_\asks(\ind{\bidi[1]\geq\tau_\asks}, \cdots, \ind{\bidi[n]\geq\tau_\asks})$,
        \item $\forall \bids \in [0,1]^{n}$, there exists $\theta_\bids \in [0,1]$ and a monotone Boolean function $g_\bids:\{0,1\}^{n}\to\{0,1\}$ such that $\alloc(\bids,\asks) = g_\bids(\ind{\askj[1]\leq\theta_\bids}, \cdots, \ind{\askj[n]\leq\theta_\bids})$.
    \end{enumerate}
\end{theorem}

\begin{theorem}
	\label{thm:deterministic_characterization_sbb}
	A deterministic allocation function $\alloc(\bids,\asks)$ for MBT can be implemented by an IC and SBB mechanism if and only if there exists $\tau \in [0,1]$ and a monotone Boolean function $f:\{0,1\}^{2n}\to\{0,1\}$ such that $
		\alloc(\bids,\asks) = 
		f(\ind{\bidi[1]\geq\tau}, \cdots, \ind{\bidi[n]\geq\tau},
		  \ind{\askj[1]\leq\tau}, \cdots, \ind{\askj[n]\leq\tau})
	$.
\end{theorem}

To give proofs to both of the theorems, we first invoke Myerson's Lemma from \cite{myerson1981optimal}.

\begin{lemma}[Myerson's Lemma \cite{myerson1981optimal}]
	\label{lemma:myerson}
	Consider a single-parameter environment with $n$ bidders where bidder $i$ has a private value $\vali \sim \disti$. Let the utility of bidder $i$ be $\utili(\bids) = \vali \cdot \alloci(\bids) - \payi(\bids)$. Here $\bids$ are the bidders' bids and $\alloci(\bids), \payi(\bids)$ are  defined by the mechanism. Then
	\begin{enumerate}[\hspace{1em}\normalfont(a)]
		\item Allocation function $\allocs$ is implementable by an IC mechanism if and only if for every $i\in [n]$, $\alloci(\bids)$ is non-decreasing in $\bidi$.
		\item If $\forall i\in [n]$, $\alloci(\bids)$ is non-decreasing in $\bidi$, then mechanism $(\allocs,\pays)$ is IC if and only if
		\begin{align*}
			\payi(\bidi,\bidsmi) = \bidi \cdot \alloci(\bidi,\bidsmi) - \int_{z=-\infty}^{\bidi} \alloci(z,\bidsmi)\d z\\
            +h_i(\bidsmi)\quad \forall i\in[n].
		\end{align*}
		Here, $h_i$ is a function unrelated to $\bidi$ for each $i\in[n]$.
	\end{enumerate}
\end{lemma}

Note that in the context of \cref{lemma:myerson}, the allocation $\alloci$ and payment $\payi$ functions for each bidder $i$ can be different, but in our MBT setting, the allocation function $\alloc$ is shared by all participants, and the payment $\pay$ and receipt $\rec$ functions are shared by all buyers and sellers respectively. This further restricts the range of feasible allocation functions. A translated version of \cref{lemma:myerson} in our setting is presented as \cref{lemma:myerson_with_two_groups} below with its proof deferred to \cref{appsub:proof_of_lemma_myerson_with_two_groups}.

\begin{lemma}
    \label{lemma:myerson_with_two_groups}
    If an allocation function $\alloc(\bids,\asks)$ for MBT can be implemented by an IC mechanism $(\alloc, \pay, \rec)$, then all of the following claims hold.
    \begin{enumerate}[\hspace{1em}\normalfont(a)]
        \item $\alloc(\bids,\asks)$ is non-decreasing in $\bidi$ for all $i\in[n]$ and non-increasing in $\askj$ for all $j\in[n]$.
        \item There exist functions $h_i$ for each $i \in [n]$ such that
        \begin{align*}
            \pay(\bidi,\bidsmi,\asks)
        = \bidi \cdot \alloc(\bidi,\bidsmi,\asks) - \int_{z=-\infty}^{\bidi} \alloc(z,\bidsmi,\asks)\d z \\
        + h_i(\bidsmi,\asks)\quad \forall i\in[n].
        \end{align*}
        \item There exist functions $h_j$ for each $j \in [n]$ such that
        \begin{align*}
            \rec(\bids,\askj,\asksmj)
        = \askj \cdot \alloc(\bids,\askj,\asksmj) + \int_{z=\askj}^{+\infty} \alloc(\bids,z,\asksmj)\d z\\
        +h_j(\bids,\asksmj)\quad \forall j\in[n].
        \end{align*}
        \item $\{(\bids,\asks) \mid \alloc(\bids,\asks) = 1\}$ is a closed set.
    \end{enumerate}
\end{lemma}


Applying \cref{lemma:myerson_with_two_groups}, we will be able to prove \cref{lemma:allocation_to_payments} which relates the allocation function to the payment and receipt functions. Its proof can be found in \cref{appsub:proof_of_lemma_allocation_to_payments}.

\begin{lemma}
    \label{lemma:allocation_to_payments}
    If an allocation function $\alloc(\bids,\asks)$ for MBT can be implemented by an IC mechanism $(\alloc, \pay, \rec)$, then for any $0\leq \alpha\leq \beta\leq 1$ and $i,j\in [n]$
    \begin{align*}
        &\textup{(a)}\quad\alloc(\alpha, \bidsmi, \asks) = \alloc(\beta, \bidsmi, \asks) \implies \pay(\alpha, \bidsmi, \asks) = \pay(\beta, \bidsmi, \asks),\\
        &\textup{(b)}\quad\alloc(\bids, \alpha, \asksmj) = \alloc(\bids, \beta, \asksmj) \implies \rec(\bids, \alpha, \asksmj) = \rec(\bids, \beta, \asksmj).
    \end{align*}
\end{lemma}



Note that for SBB mechanisms, $\pay(\bids, \asks) = \rec(\bids, \asks)$. By iteratively applying \cref{lemma:allocation_to_payments} to each applicable coordinate, we can see the following \cref{coro:allocation_to_payments} holds.

\begin{corollary}
    \label{coro:allocation_to_payments}
    If an allocation function $\alloc(\bids,\asks)$ for MBT can be implemented by an IC mechanism $(\alloc, \pay, \rec)$, then for any $\bids^{(0)} \preceq \bids^{(1)}, \asks^{(0)} \succeq \asks^{(1)}$,
    \begin{align*}
        &\textup{(a)}\quad \alloc(\bids^{(0)}, \asks) = \alloc(\bids^{(1)}, \asks) \implies 
        \pay(\bids^{(0)}, \asks) = \pay(\bids^{(1)}, \asks),\\
        &\textup{(b)}\quad \alloc(\bids, \asks^{(0)}) = \alloc(\bids, \asks^{(1)}) \implies 
        \rec(\bids, \asks^{(0)}) = \rec(\bids, \asks^{(1)}).
    \end{align*}
    Moreover, if $\alloc(\bids,\asks)$ for MBT can be implemented by an IC and SBB mechanism,
    \begin{align*}
        &\textup{(c)}\quad \alloc(\bids^{(0)}, \asks^{(0)}) = \alloc(\bids^{(1)}, \asks^{(1)}) \implies \\
        &\pay(\bids^{(0)}, \asks^{(0)}) = \pay(\bids^{(1)}, \asks^{(1)}) = \rec(\bids^{(0)}, \asks^{(0)}) = \rec(\bids^{(1)}, \asks^{(1)}).
    \end{align*}
    Here $\preceq$ and $\succeq$ mean componentwise $\leq$ and $\geq$ respectively.
\end{corollary}

Let $\vec 0, \vec 1$ be vectors of $n$ zeros and $n$ ones respectively. Next, we will present two lemmas connecting back to \cref{thm:deterministic_characterization} and \cref{thm:deterministic_characterization_sbb} with their proofs deferred to \cref{appsub:proof_of_lemma_necessity,appsub:proof_of_lemma_necessity_sbb}.

\begin{lemma}
    \label{lemma:necessity}
    If a deterministic allocation function $\alloc(\bids,\asks)$ for MBT can be implemented by an IC mechanism, then the following statements hold.
    \begin{enumerate}[\hspace{0em}\normalfont(a)]
        \item For any $\asks\in[0,1]^n$, let $\tau_\asks = \pay(\vec 1, \asks) - \pay(\vec 0, \asks)$, then
            \begin{equation*}
                \ind{\alpha \geq \tau_\asks} = \ind{\beta \geq \tau_\asks}
                \implies
                \alloc(\alpha, \bidsmi, \asks) = \alloc(\beta, \bidsmi, \asks).
            \end{equation*}
        \item For any $\bids\in[0,1]^n$, let $\theta_\bids = \rec(\bids, \vec 0) - \rec(\bids, \vec 1)$, then
            \begin{equation*}
                \ind{\alpha \leq \theta_\bids} = \ind{\beta \leq \theta_\bids}
                \implies
                \alloc(\bids, \alpha, \asksmj) = \alloc(\bids, \beta, \asksmj).
            \end{equation*}
    \end{enumerate}
\end{lemma}

There is also a version of \cref{lemma:necessity} for IC and SBB mechanisms with an almost identical proof.

\begin{lemma}
    \label{lemma:necessity_sbb}
    If a deterministic allocation function $\alloc(\bids,\asks)$ for MBT can be implemented by an IC and SBB mechanism $(\alloc, \pay, \rec)$, then let $\tau = \pay(\vec 1, \vec 0) - \pay(\vec 0, \vec 1)$, $\forall 0\leq \alpha\leq \beta\leq 1$,
    \begin{align*}
        &\textup{(a)}\ 
        \ind{\alpha \geq \tau} = \ind{\beta \geq \tau}
        \implies
        \alloc(\alpha, \bidsmi, \asks) = \alloc(\beta, \bidsmi, \asks),\\
        &\textup{(b)}\ 
        \ind{\alpha \leq \tau} = \ind{\beta \leq \tau}
        \implies
        \alloc(\bids, \alpha, \asksmj) = \alloc(\bids, \beta, \asksmj).
    \end{align*}
\end{lemma}

With all of the above lemmas, we are able to prove \cref{thm:deterministic_characterization} and \cref{thm:deterministic_characterization_sbb}. In fact, the necessity of \cref{thm:deterministic_characterization} (resp. \cref{thm:deterministic_characterization_sbb}) is already implied by \cref{lemma:necessity} (resp. \cref{lemma:necessity_sbb}). The proofs of the sufficiency of these theorems can be found in \cref{appsub:proof_of_thm_deterministic_characterization,appsub:proof_of_thm_deterministic_characterization_sbb}.

\subsection{When Expected Value is Greater than Expected Cost: A Simple Efficient Mechanism}
\label{subsec:deterministic_positive}

\cref{thm:deterministic_characterization,thm:deterministic_characterization_sbb} provide a characterization of the design space. Under this design space, we will first show a simple mechanism that is IC, SBB, IR in the limit and efficient in the limit when the expected value is greater than the expected cost, i.e., $\Ex[\val\sim\bdist]{\val}>\Ex[\cost\sim\sdist]{\cost}$.
\begin{algorithm}[htbp]
    \caption{Forced-Trade Mechanism}
    \label{alg1}

    Trade if and only if $\Ex[\val\sim\bdist]{\val} > \Ex[\cost\sim\sdist]{\cost}$, with both payment and receipt being $0.5(\Ex[\val\sim\bdist]{\val}+\Ex[\cost\sim\sdist]{\cost})$. I.e.,
    \begin{align*}
        \alloc(\bids,\asks) &= \ind{\Ex[\val\sim\bdist]{\val} > \Ex[\cost\sim\sdist]{\cost}},\\
        \pay(\bids,\asks) &= \rec(\bids,\asks) = 0.5(\Ex[\val\sim\bdist]{\val}+\Ex[\cost\sim\sdist]{\cost})\cdot \alloc(\bids,\asks).
    \end{align*}
\end{algorithm}

We show \cref{alg1} achieves all desiderata in the limit.

\begin{theorem}
    \label{thm:forced_trade} 
    \cref{alg1} is IC and SBB. When $\Ex[\val\sim\bdist]{\val}>\Ex[\cost\sim\sdist]{\cost}$, as $n\to\infty$, \cref{alg1} is IR with probability $1-e^{-\Omega(n)}$, and $\frac{\ALG}{\FB}=1-e^{-\Omega(n)}$ where $\ALG$ is its GFT.
\end{theorem}

The proof of \cref{thm:forced_trade} can be found in \cref{appsub:proof_of_thm_forced_trade}.

\subsection{When Expected Value is No Greater than Expected Cost: No Efficient Mechanisms}
\label{subsec:deterministic_hardness}

\cref{thm:forced_trade} shows that when the expected value is greater than the expected cost, the extremely simple \cref{alg1} achieves all desiderata in the limit. In our next \cref{thm:deterministic_hardness}, we will additionally show that when this is not the case, no mechanisms can be both IC and efficient in the limit.

\begin{theorem}
    \label{thm:deterministic_hardness}
    When $\Ex[\val\sim\bdist]{\val}\leq \Ex[\cost\sim\sdist]{\cost}$, no IC mechanisms for MBT can be efficient in the limit.
\end{theorem}

The proof of \cref{thm:deterministic_hardness} is built on our characterization (\cref{thm:deterministic_characterization}) and deferred to \cref{appsub:proof_of_thm_deterministic_hardness}. By \cref{thm:deterministic_hardness}, when the expected value is no greater than the expected cost, the multiplicative approximation ratio of any IC deterministic mechanism is $1-\Omega(1)$, i.e., no deterministic IC mechanism can be efficient in the limit. However, recall that \cref{lemma:fb_values} shows that in this case, the first best is $o(n)$. This implies that {\em additively}, \cref{alg1} does not lose much compared to the first best even if it does not trade at all.

    


\begin{table*}[htbp]
	\centering
	\begin{tabular}{|c|c|c|c|c|c|c|c|}
		\hline
		\multicolumn{2}{|c|}{$(\mu_\bdist,\mu_\sdist)$} & \multicolumn{2}{c|}{$(0.6, 0.4)$} & \multicolumn{2}{c|}{$(0.55, 0.45)$} & \multicolumn{2}{c|}{$(0.51, 0.49)$}  \\
		\hline
		Distribution & $n$ & IR & Efficiency &  IR & Efficiency & IR & Efficiency \\
		\hline
		\multirow{3}{*}{Normal} & 5 & 0.804033 & 0.992350 & 0.539705 & 0.899567 & 0.302261 & 0.359552 \\
		\cline{2-8}
		& 100 & 1.000000 & 1.000000 & 0.995614 & 0.999999 & 0.510591 & 0.870741 \\
		\cline{2-8}
		& 10000 & 1.000000 & 1.000000 & 1.000000 & 1.000000 & 1.000000 & 1.000000 \\
		\hline
		\multirow{3}{*}{Uniform} & 5 & 0.684952 & 0.972358 & 0.464824 & 0.823521 & 0.288943 & 0.290061 \\
		\cline{2-8}
		& 100 & 0.999992 & 1.000000 & 0.969823 & 0.999900 & 0.445284 & 0.788148 \\
		\cline{2-8}
		& 10000 & 1.000000 & 1.000000 & 1.000000 & 1.000000 & 0.999989 & 1.000000 \\
		\hline
		\multirow{3}{*}{Bernoulli} & 5 & 0.465557 & 0.809241 & 0.351963 & 0.558944 & 0.268607 & 0.150281 \\
		\cline{2-8}
		& 100 & 0.966587 & 0.999779 & 0.749410 & 0.975987 & 0.382188 & 0.512118 \\
		\cline{2-8}
		& 10000 & 1.000000 & 1.000000 & 1.000000 & 1.000000 & 0.955986 & 0.999752 \\
		\hline
		\multirow{3}{*}{Mixed} & 5 & 0.566107 & 0.909725 & 0.401324 & 0.694859 & 0.277582 & 0.208867 \\
		\cline{2-8}
		& 100 & 0.997254 & 0.999999 & 0.870417 & 0.997094 & 0.382296 & 0.655883 \\
		\cline{2-8}
		& 10000 & 1.000000 & 1.000000 & 1.000000 & 1.000000 & 0.997250 & 1.000000 \\
		\hline
	\end{tabular}
	\caption{Performance of \cref{alg1} under various conditions. IR is the empirical probability of \cref{alg1} being IR and efficiency is the empirical expected $\GFT/\FB$. All data are averaged over $10^6$ runs.}
	\label{table:main_experiments}
\end{table*}

\section{Randomized Mechanisms}
\label{sec:randomized_mechanisms}

In this section, we move on to consider twice continuously differentiable randomized mechanisms for multiplayer bilateral trade (MBT). Namely, the allocation function $\alloc(\bids,\asks)\in[0,1]$ and $\alloc$ is twice continuously differentiable.  Note that \cref{lemma:fb_values} still holds in this context. Therefore, we will take a similar approach with \cref{sec:deterministic_mechanisms}. We will still first characterize the set of allocation functions that are implementable by an IC mechanism in \cref{subsec:randomized_characterization}. When the expected value is greater than the expected cost, \cref{alg1} in \cref{subsec:deterministic_positive} is still applicable as the allocation function of \cref{alg1} is also twice continuously differentiable. And when the expected value is not greater than the expected cost, we show in \cref{subsec:randomized_hardness} that no twice continuously differentiable randomized mechanism can achieve all the desiderata, similarly to \cref{sec:deterministic_mechanisms}. In other words, completely smooth randomness cannot help in MBT. Surprisingly, our negative result is stronger than \cref{thm:deterministic_hardness}.

\subsection{Characterization of Incentive Compatible Mechanisms}
\label{subsec:randomized_characterization}

We characterize the set of all twice continuously differentiable randomized allocation functions that are implementable by an IC mechanism in the following \cref{thm:randomized_characterization}.

From a general perspective, our characterization shows that such mechanisms must be separable across different agents, which means that fixing the asks (resp.\ bids) of all sellers (resp.\ buyers), for each buyer (resp.\ seller), there is a component of the allocation function that depends only on the bid of that buyer (resp.\ seller), and the overall allocation function is the sum of all these components

\begin{theorem}
	\label{thm:randomized_characterization}
	A twice continuously differentiable randomized allocation function $\alloc(\bids,\asks)$ for MBT can be implemented by an IC mechanism if and only if both of the following statements hold.
    \begin{enumerate}[\hspace{0em}\normalfont(a)]
        \item For all $\asks \in [0,1]^{n}$, there exist $n$ non-decreasing differentiable functions $f_{\asks,i}:[0,1]\to\R,\forall i \in [n]$, such that $\alloc(\bids,\asks) = f_{\asks,1}(\bidi[1]) + f_{\asks,2}(\bidi[2]) + \cdots + f_{\asks,n}(\bidi[n])$.
        \item For all $\bids \in [0,1]^{n}$, there exist $n$ non-increasing differentiable functions $g_{\bids,j}:[0,1]\to\R,\forall j \in [n]$, such that $\alloc(\bids,\asks) = g_{\bids,1}(\askj[1]) + g_{\bids,2}(\askj[2]) + \cdots + g_{\bids,n}(\askj[n])$.
    \end{enumerate}
\end{theorem}

Below, we will first provide two lemmas. \cref{thm:randomized_characterization} can then be proved by combining these two lemmas.

\begin{lemma}
    \label{lemma:calculus}
    For a twice continuously differentiable function $f(x_1,x_2,\dots,x_n):\R^n\to\R$, the following statements (i) and (ii) are equivalent. 
    \begin{enumerate}[\hspace{0em}\normalfont(i)]
        \item There exist $n$ differentiable functions $f_i:\R\to\R$, that $f(x_1,x_2,\dots,x_n) = \sum_{i=1}^{n}f_i(x_i)$.
        \item For any $i,j\in[n], i\ne j$, $\frac{\partial^2 f}{\partial x_i\partial x_j}(x_1,x_2,\dots,x_n) = 0$.
    \end{enumerate}
\end{lemma}



\begin{lemma}
    \label{lemma:independent}
    If a twice continuously differentiable randomized allocation function $\alloc(\bids,\asks)$ for MBT can be implemented by an IC mechanism, then the followings hold.
    \begin{enumerate}[\hspace{0em}\normalfont(a)]
        \item Fix $\asks\in[0,1]^n$,
        \begin{equation*}
            \frac{\partial^2 \alloc}{\partial \bidi[i1]\partial \bidi[i2]}(\bids,\asks) = 0, \forall i_1,i_2\in [n], i_1\ne i_2.
        \end{equation*}
        \item Fix $\bids\in[0,1]^n$,
        \begin{equation*}
            \frac{\partial^2 \alloc}{\partial \askj[j1]\partial \askj[j2]}(\bids,\asks) = 0, \forall j_1,j_2\in [n], j_1\ne j_2.
        \end{equation*}
    \end{enumerate}
\end{lemma}

With \cref{lemma:calculus,lemma:independent}, we are ready to show \cref{thm:randomized_characterization}. The detailed proofs of \cref{lemma:calculus,lemma:independent,thm:randomized_characterization} are deferred to \cref{appsub:proof_of_lemma_calculus,appsub:proof_of_lemma_independent,appsub:proof_of_thm_randomized_characterization}.

\subsection{When Expected Value is No Greater than Expected Cost: No Efficient Mechanisms}
\label{subsec:randomized_hardness}

At first glance, one might think twice continuously differentiable randomized mechanisms should be more powerful than deterministic mechanisms because the choice of allocation function seems much wider. However, we show in the following \cref{thm:randomized_hardness} that on the contrary, these mechanisms are not even capable of being a constant approximation of $\FB$ in the limit.

\begin{theorem}
    \label{thm:randomized_hardness}
    When $\Ex[\val\sim\bdist]{\val}\leq \Ex[\cost\sim\sdist]{\cost}$, no IC twice continuously differentiable mechanisms for MBT can have a GFT that is a constant approximation of $\FB$ in the limit.
\end{theorem}

The proof of \cref{thm:randomized_hardness} can be found in \cref{appsub:proof_of_thm_randomized_hardness}.

\section{Experiments}
\label{sec:experiments}

\begin{figure*}[htbp]
	\centering
	\begin{subfigure}{0.32\textwidth}
	  \centering
	  \includegraphics[width=\linewidth]{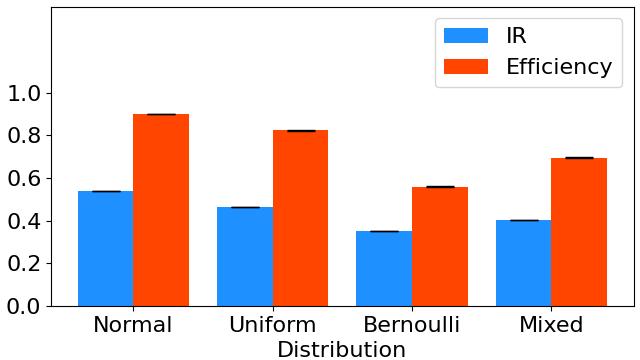}
	  \caption{$n = 5$}
	  \label{subfigure:n=5}
	\end{subfigure}
	\hfill
	\begin{subfigure}{0.32\textwidth}
	  \centering
	  \includegraphics[width=\linewidth]{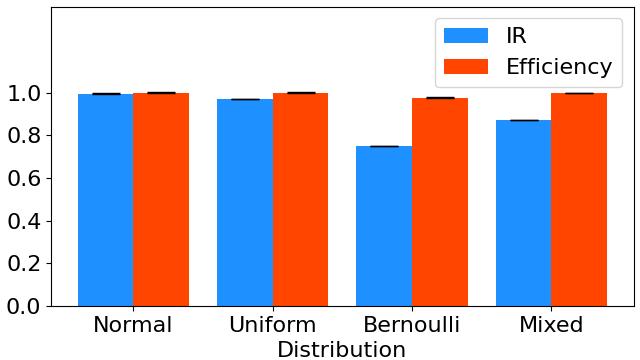}
	  \caption{$n = 100$}
	  \label{subfigure:n=100}
	\end{subfigure}
	\hfill
	\begin{subfigure}{0.32\textwidth}
	  \centering
	  \includegraphics[width=\linewidth]{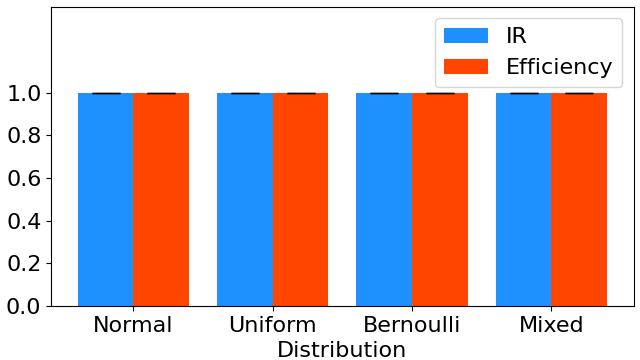}
	  \caption{$n = 10000$}
	  \label{subfigure:n=10000}
	\end{subfigure}
	\caption{Performance of \cref{alg1} when $(\mu_\bdist,\mu_\sdist)=(0.55,0.45)$ and $n=5,100,10000$. Error bars indicate standard error of mean. The empirical probability of being IR and the empirical efficiency converge to $1.0$ as $n$ increases.  \cref{alg1} has a good performance for reasonably large $n$. Errors are negligible due to the large number of samples.}
	\label{fig:varyn}
\end{figure*}

\begin{figure*}[htbp]
	\centering
	\begin{subfigure}{0.32\textwidth}
	  \centering
	  \includegraphics[width=\linewidth]{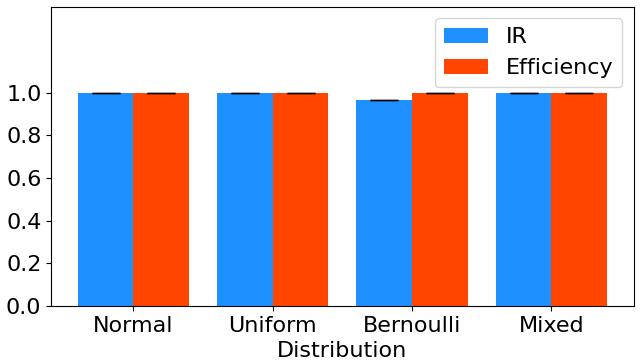}
	  \caption{$(\mu_\bdist,\mu_\sdist)=(0.6,0.4)$}
	  \label{subfigure:de=0.2}
	\end{subfigure}
	\hfill
	\begin{subfigure}{0.32\textwidth}
	  \centering
	  \includegraphics[width=\linewidth]{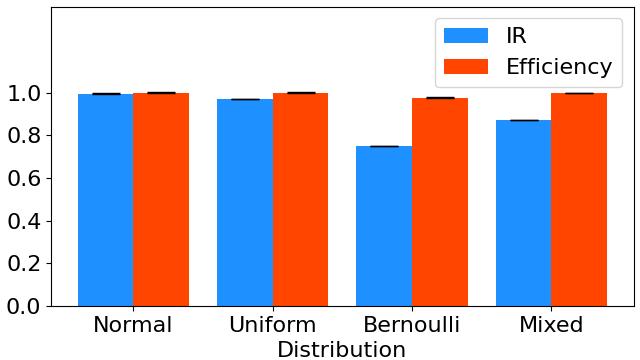}
	  \caption{$(\mu_\bdist,\mu_\sdist)=(0.55,0.45)$}
	  \label{subfigure:de=0.1}
	\end{subfigure}
	\hfill
	\begin{subfigure}{0.32\textwidth}
	  \centering
	  \includegraphics[width=\linewidth]{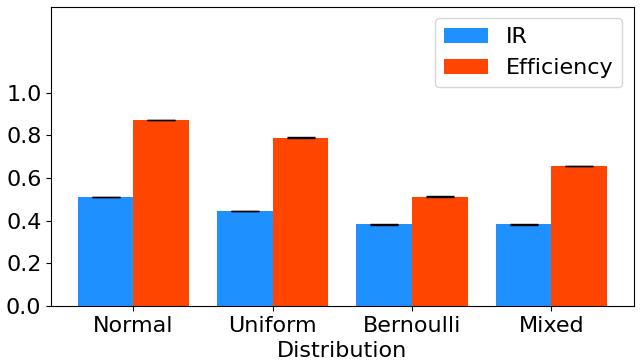}
	  \caption{$(\mu_\bdist,\mu_\sdist)=(0.51,0.49)$}
	  \label{subfigure:de=0.02}
	\end{subfigure}
	\caption{Performance of \cref{alg1} when $(\mu_\bdist,\mu_\sdist)=(0.6,0.4),(0.55,0.45),(0.51,0.49)$ and $n=100$. Error bars indicate standard error of mean. The empirical probability of being IR and the empirical efficiency decrease as $\mu_\bdist-\mu_\sdist$ gets smaller. When $\mu_\bdist$ and $\mu_\sdist$ are not too close, $n\geq 100$ is enough to guarantee good performance of \cref{alg1}. Errors are negligible due to the large number of samples.}
	\label{fig:varye}
\end{figure*}

In the previous sections, we have shown that the simple \cref{alg1} achieves all desiderata in the limit asymptotically when $\Ex[\val\sim\bdist]{\val} > \Ex[\cost\sim\sdist]{\cost}$ and no mechanisms can do so when $\Ex[\val\sim\bdist]{\val}\leq \Ex[\cost\sim\sdist]{\cost}$. In this section, we study the non-asymptotic performance of \cref{alg1} by running experiments on various generated datasets. Only simulated data without real individuals are involved in the experiments. As \cref{alg1} is always IC and SBB, we focus on studying the empirical probability that \cref{alg1} is IR and its empirical efficiency on datasets of practically relevant sizes.

\textbf{Experiment setup.} We implemented \cref{alg1} in C++. The source code can be found at \url{https://github.com/YixuanEvenXu/bilateral-trade}. The datasets we used were generated as follows. The distributions $\bdist$ and $\sdist$ are both supported on $[0,1]$. In every dataset, $\bdist$ and $\sdist$ belong to one of the $4$ families of distributions given below and were truncated to the specified support whenever necessary.
\begin{enumerate}[\hspace{1.5em}(1)]
    \item Normal: $\bdist,\sdist$ are normal distributions $\mathcal N(\mu,\sigma^2)$ where $\mu = \mu_F$ or $\mu_G$ and $\sigma = 0.2$.
    \item Uniform: $\bdist,\sdist$ are uniform distributions $\mathcal U[\mu - r, \mu + r]$ where $\mu = \mu_F$ or $\mu_G$ and $r = 0.4$.
    \item Bernoulli: $\bdist,\sdist$ are Bernoulli distributions $\mathrm{Ber}(\mu)$ where $\mu = \mu_F$ or $\mu_G$.
    \item Mixed: $\bdist,\sdist$ are mixed distributions with $\frac{1}{3}$ probability of (1), (2), and (3) each.
\end{enumerate}

For each type of distribution above, we performed various tests with $n = \{5, 100, 10000\}$ and $(\mu_\bdist,\mu_\sdist) = \{(0.6, 0.4), (0.55, 0.45), (0.51, 0.49)\}$. The complete experimental results are shown in \cref{table:main_experiments}. All of the data are averaged over $10^6$ runs. We also highlight a few results in \cref{fig:varye,fig:varyn} to make it easier to interpret the results.

\paragraph{Comparing when $n=5,100,$ and $10000$.} As shown in \cref{table:main_experiments,fig:varyn}, \cref{alg1}'s empirical probability of being IR and empirical efficiency generally increase as $n$ scales up. This is consistent with the asymptotic guarantees of \cref{alg1} in \cref{thm:forced_trade}. Moreover, as long as $\mu_{\bdist}-\mu_{\sdist}$ is not too small, $n$ does not have to be very large to ensure a high probability of IR and efficiency.

\paragraph{Comparing when $(\mu_\bdist,\mu_\sdist)=(0.6,0.4),(0.55,0.45),$ and $(0.51,0.49).$} As shown in \cref{table:main_experiments,fig:varye}, Algorithm \ref{alg1}'s empirical probability of being IR and empirical efficiency generally decrease as $\mu_\bdist-\mu_\sdist$ goes down. Note that \cref{thm:deterministic_hardness,thm:randomized_hardness} show that it is impossible to be efficient in the limit when $\mu_\bdist\leq \mu_\sdist$. The experiments imply that the closer an instance is to $\mu_\bdist\leq \mu_\sdist$, the harder it is to solve it well. Moreover, even if $\mu_\bdist-\mu_\sdist$ is extremely small, i.e., $(\mu_{\bdist},\mu_{\sdist})=(0.51,0.49)$, with a sufficiently large $n$, say $n\geq 10000$, \cref{alg1}'s empirical probability of IR and empirical efficiency can still be very close to $1.0$.

\paragraph{Comparing different types of distributions.} For each set of $(n,\mu_{\bdist},\mu_{\sdist})$, we can observe that the performance of \cref{alg1} is generally better on normal and uniform distributions, and worse on Bernoulli and mixed distributions. This is because the variances of normal and uniform distributions are lower than those of Bernoulli and mixed distributions, suggesting that the performance of \cref{alg1} is better if the distributions $\bdist,\sdist$ have lower variances.

\section{Discussions}
\label{sec:discussions}

In this paper, we conducted a thorough investigation of bilateral trade when there are multiple self-interested agents on both sides of the trade.
Our results suggest that we can achieve all desiderata in the limit, as long as buyers value the item more than sellers in expectation.
Alternatively, when sellers value the item (weakly) more, efficiency is impossible with feasible deterministic mechanisms, and smooth randomness does not help, either. On the other hand, we showed that in these negative cases, even the first-best cannot obtain much value, so not much value is lost.
These results extend and complement existing results in classical settings of bilateral trade.
In addition, our characterizations of feasible mechanisms might be of independent interest in other mechanism design problems with shared allocation and payment.

While the setting we study is motivated by real-world applications, there are other settings that are also practically sensible.
For example, in certain trading scenarios, it might be possible to exclude a subset of agents from the trade. Such settings are reminiscent of mechanism design for cost sharing~\citep{moulin1992serial}.
One might also consider relaxed notions of incentive compatibility, or personalized payments, which would greatly expand the design space and allow for more efficient mechanisms.
Another practical consideration is prior-independence: Can we design a mechanism that achieves efficiency independent of the value and cost distributions?
Given our characterization, it is not hard to show that this cannot be done in the setting considered in this paper, but there might be possibilities if one, for example, considers a relaxed notion of incentive compatibility.

\newpage

\section*{Acknowledgements}

HZ was supported by the National Science Foundation under Grant No. DMS-1928930 and by the Alfred P. Sloan Foundation under grant G-2021-16778, while HZ was in residence at the Simons Laufer Mathematical Sciences Institute (formerly MSRI) in Berkeley, California, during the Fall 2023 semester.  HZ and VC also thank the Cooperative AI
Foundation, Polaris Ventures (formerly the Center for Emerging Risk Research), and Jaan Tallinn's
donor-advised fund at Founders Pledge for financial support.

\bibliography{ref}

\begin{thebibliography}{25}
\providecommand{\natexlab}[1]{#1}

\bibitem[{Babaioff et~al.(2018)Babaioff, Cai, Gonczarowski, and
  Zhao}]{babaioff2018best}
Babaioff, M.; Cai, Y.; Gonczarowski, Y.~A.; and Zhao, M. 2018.
\newblock The Best of Both Worlds: Asymptotically Efficient Mechanisms with a
  Guarantee on the Expected Gains-From-Trade.
\newblock In \emph{Proceedings of the 2018 ACM Conference on Economics and
  Computation}, 373--373.

\bibitem[{Babaioff, Goldner, and Gonczarowski(2020)}]{babaioff2020bulow}
Babaioff, M.; Goldner, K.; and Gonczarowski, Y.~A. 2020.
\newblock Bulow-klemperer-style results for welfare maximization in two-sided
  markets.
\newblock In \emph{Proceedings of the Fourteenth Annual ACM-SIAM Symposium on
  Discrete Algorithms}, 2452--2471. SIAM.

\bibitem[{Balseiro et~al.(2019)Balseiro, Mirrokni, Leme, and
  Zuo}]{doi:10.1137/1.9781611975482.11}
Balseiro, S.~R.; Mirrokni, V.; Leme, R.~P.; and Zuo, S. 2019.
\newblock Dynamic Double Auctions: Towards First Best.
\newblock In \emph{Proceedings of the 2019 Annual ACM-SIAM Symposium on
  Discrete Algorithms}, 157--172.

\bibitem[{Balseiro et~al.(2021)Balseiro, Mirrokni, Leme, and
  Zuo}]{balseiro2021non}
Balseiro, S.~R.; Mirrokni, V.; Leme, R.~P.; and Zuo, S. 2021.
\newblock Non-excludable dynamic mechanism design.
\newblock In \emph{Proceedings of the 2021 ACM-SIAM Symposium on Discrete
  Algorithms (SODA)}, 1357--1373. SIAM.

\bibitem[{Blumrosen and Mizrahi(2016)}]{blumrosen2016approximating}
Blumrosen, L.; and Mizrahi, Y. 2016.
\newblock Approximating gains-from-trade in bilateral trading.
\newblock In \emph{Web and Internet Economics: 12th International Conference,
  WINE 2016, Montreal, Canada, December 11-14, 2016, Proceedings 12}, 400--413.
  Springer.

\bibitem[{Brustle et~al.(2017)Brustle, Cai, Wu, and
  Zhao}]{brustle2017approximating}
Brustle, J.; Cai, Y.; Wu, F.; and Zhao, M. 2017.
\newblock Approximating gains from trade in two-sided markets via simple
  mechanisms.
\newblock In \emph{Proceedings of the 2017 ACM Conference on Economics and
  Computation}, 589--590.

\bibitem[{Buchfuhrer, Schapira, and Singer(2010)}]{buchfuhrer2010computation}
Buchfuhrer, D.; Schapira, M.; and Singer, Y. 2010.
\newblock Computation and incentives in combinatorial public projects.
\newblock In \emph{Proceedings of the 11th ACM conference on Electronic
  commerce}, 33--42.

\bibitem[{Cai et~al.(2021)Cai, Goldner, Ma, and Zhao}]{cai2021multi}
Cai, Y.; Goldner, K.; Ma, S.; and Zhao, M. 2021.
\newblock On multi-dimensional gains from trade maximization.
\newblock In \emph{Proceedings of the 2021 ACM-SIAM Symposium on Discrete
  Algorithms (SODA)}, 1079--1098. SIAM.

\bibitem[{Colini-Baldeschi et~al.(2017)Colini-Baldeschi, Goldberg, de~Keijzer,
  Leonardi, and Turchetta}]{colini2017fixed}
Colini-Baldeschi, R.; Goldberg, P.; de~Keijzer, B.; Leonardi, S.; and
  Turchetta, S. 2017.
\newblock Fixed price approximability of the optimal gain from trade.
\newblock In \emph{Web and Internet Economics: 13th International Conference,
  WINE 2017, Bangalore, India, December 17--20, 2017, Proceedings 13},
  146--160. Springer.

\bibitem[{Colini-Baldeschi et~al.(2020)Colini-Baldeschi, Goldberg, Keijzer,
  Leonardi, Roughgarden, and Turchetta}]{colini2020approximately}
Colini-Baldeschi, R.; Goldberg, P.~W.; Keijzer, B.~d.; Leonardi, S.;
  Roughgarden, T.; and Turchetta, S. 2020.
\newblock Approximately efficient two-sided combinatorial auctions.
\newblock \emph{ACM Transactions on Economics and Computation (TEAC)}, 8(1):
  1--29.

\bibitem[{Colini-Baldeschi et~al.(2016)Colini-Baldeschi, Keijzer, Leonardi, and
  Turchetta}]{colini2016approximately}
Colini-Baldeschi, R.; Keijzer, B.~d.; Leonardi, S.; and Turchetta, S. 2016.
\newblock Approximately efficient double auctions with strong budget balance.
\newblock In \emph{Proceedings of the twenty-seventh annual ACM-SIAM symposium
  on Discrete algorithms}, 1424--1443. SIAM.

\bibitem[{Deng et~al.(2022)Deng, Mao, Sivan, and Wang}]{deng2022approximately}
Deng, Y.; Mao, J.; Sivan, B.; and Wang, K. 2022.
\newblock Approximately efficient bilateral trade.
\newblock In \emph{Proceedings of the 54th Annual ACM SIGACT Symposium on
  Theory of Computing}, 718--721.

\bibitem[{Dobzinski(2011)}]{dobzinski2011impossibility}
Dobzinski, S. 2011.
\newblock An impossibility result for truthful combinatorial auctions with
  submodular valuations.
\newblock In \emph{Proceedings of the forty-third annual ACM symposium on
  Theory of computing}, 139--148.

\bibitem[{Dughmi, Roughgarden, and Yan(2016)}]{dughmi2016optimal}
Dughmi, S.; Roughgarden, T.; and Yan, Q. 2016.
\newblock Optimal mechanisms for combinatorial auctions and combinatorial
  public projects via convex rounding.
\newblock \emph{Journal of the ACM (JACM)}, 63(4): 1--33.

\bibitem[{Fei(2022)}]{fei2022improved}
Fei, Y. 2022.
\newblock Improved approximation to first-best gains-from-trade.
\newblock In \emph{Web and Internet Economics: 18th International Conference,
  WINE 2022, Troy, NY, USA, December 12--15, 2022, Proceedings}, 204--218.
  Springer.

\bibitem[{Hardin(1968)}]{10.2307/1724745}
Hardin, G. 1968.
\newblock The Tragedy of the Commons.
\newblock \emph{Science}.

\bibitem[{McAfee(1992)}]{mcafee1992dominant}
McAfee, R.~P. 1992.
\newblock A dominant strategy double auction.
\newblock \emph{Journal of economic Theory}, 56(2): 434--450.

\bibitem[{McAfee(2008)}]{mcafee2008gains}
McAfee, R.~P. 2008.
\newblock The gains from trade under fixed price mechanisms.
\newblock \emph{Applied Economics Research Bulletin}, 1(1): 1--10.

\bibitem[{Moulin and Shenker(1992)}]{moulin1992serial}
Moulin, H.; and Shenker, S. 1992.
\newblock Serial cost sharing.
\newblock \emph{Econometrica: Journal of the Econometric Society}, 1009--1037.

\bibitem[{Mugrave(1959)}]{mugrave1959theory}
Mugrave, R.~A. 1959.
\newblock The theory of public finance: A study in public economy.
\newblock \emph{London and New York: McGraw-Hill}.

\bibitem[{Myerson(1981)}]{myerson1981optimal}
Myerson, R.~B. 1981.
\newblock Optimal auction design.
\newblock \emph{Mathematics of operations research}, 6(1): 58--73.

\bibitem[{Myerson and Satterthwaite(1983)}]{myerson1983efficient}
Myerson, R.~B.; and Satterthwaite, M.~A. 1983.
\newblock Efficient mechanisms for bilateral trading.
\newblock \emph{Journal of economic theory}, 29(2): 265--281.

\bibitem[{Ostrom(1990)}]{ostrom1990governing}
Ostrom, E. 1990.
\newblock \emph{Governing the commons: The evolution of institutions for
  collective action}.
\newblock Cambridge university press.

\bibitem[{Papadimitriou, Schapira, and
  Singer(2008)}]{papadimitriou2008hardness}
Papadimitriou, C.; Schapira, M.; and Singer, Y. 2008.
\newblock On the hardness of being truthful.
\newblock In \emph{2008 49th Annual IEEE Symposium on Foundations of Computer
  Science}, 250--259. IEEE.

\bibitem[{Samuelson(1954)}]{samuelson1954pure}
Samuelson, P.~A. 1954.
\newblock The pure theory of public expenditure.
\newblock \emph{The review of economics and statistics}, 387--389.

\end{thebibliography}

\newpage
\appendix

\onecolumn

\section{Missing Proofs in Section \ref{sec:deterministic_mechanisms}}
\label{app:deterministic_mechanisms}

\subsection{Proof of \cref{lemma:fb_values}}
\label{appsub:proof_of_lemma_fb_values}

\begin{proofof}{\cref{lemma:fb_values}}
    Let $\hat \mu_\val = \frac{1}{n} \sum_{i} \vali$, $\hat \mu_\cost = \frac{1}{n} \sum_{j} \costj$,
    and let the mean and variance of $\bdist,\sdist$ be $\mu_\val,\sigma^2_\val$, $\mu_\cost,\sigma^2_\cost$, respectively.
    Compute that $\FB = n \cdot \Ex[\vals \sim \bdist^n, \costs \sim \sdist^n] {(\hat \mu_\val - \hat \mu_\cost) \cdot \ind{\hat \mu_\val \geq \hat \mu_\cost}}$. Note that for $i\in [n]$ (resp. $j\in [n]$), $\vali$ (resp. $\costj$) are i.i.d$.$ random variables with mean $\mu = \mu_\val$ (resp. $\mu_\cost$) and variance $\sigma^2 = \sigma^2_\val$ (resp. $\sigma^2_\cost$). Using the central limit theorem, $\sqrt{n}(\hat \mu_\val - \mu_\val)$ (resp. $\sqrt{n}(\hat \mu_\cost - \mu_\cost)$) converges in distribution to a normal distribution $\N(0,\sigma^2_\val)$ (resp. $\N(0,\sigma^2_\cost)$).
    
    When $\mu_\val \leq \mu_\cost$,
    \begin{align*}
        &n\cdot (\hat \mu_\val - \hat \mu_\cost) \cdot \ind{\hat \mu_\val \geq \hat \mu_\cost}\\
        \leq &\ \sqrt{n}\cdot \InParentheses{\sqrt{n}(\hat \mu_\val - \mu_\val) - \sqrt{n}(\hat \mu_\cost - \mu_\val)} \cdot \ind{\sqrt{n}(\hat \mu_\val - \mu_\val) \geq \sqrt{n}(\hat \mu_\cost - \mu_\val)}.
    \end{align*}
    Thus,
    \begin{align*}
        \FB \leq (1 + o(1))\sqrt{n}\cdot \Ex[x\sim \N(0,\sigma^2_\val),y\sim \N(0,\sigma^2_\cost)]{(x-y) \cdot \ind{x\geq y}} = O(\sqrt{n}).
    \end{align*}

    On the other hand, when $\mu_\val > \mu_\cost$, 
    \begin{align*}
        \FB &= n \cdot \Ex[\vals \sim \bdist^n, \costs \sim \sdist^n] {(\hat \mu_\val - \hat \mu_\cost) \cdot \ind{\hat \mu_\val \geq \hat \mu_\cost}}\\
        &\geq n \cdot \Ex[\vals \sim \bdist^n, \costs \sim \sdist^n] {(\hat \mu_\val - \hat \mu_\cost)} \\
        &= n(\mu_\val - \mu_\cost) = \Omega(n).
    \end{align*}
    This concludes the proof of \cref{lemma:fb_values}.
\end{proofof}

\subsection{Proof of \cref{lemma:myerson_with_two_groups}}
\label{appsub:proof_of_lemma_myerson_with_two_groups}

\begin{proofof}{\cref{lemma:myerson_with_two_groups}}
    Recall that each buyer $i$ has a utility $\utili(\bids,\asks) = \vali\cdot\alloc(\bids,\asks) -\pay(\bids,\asks)$ and each seller $j$ has a utility $\utilj(\bids,\asks) = \rec(\bids,\asks) - \costj\cdot\alloc(\bids,\asks) = (-\costj)\cdot\alloc(\bids,\asks) - (-\rec(\bids,\asks))$. Consider buyer $i$ as a bidder with private value $\vali$ and seller $j$ as a bidder with private value $-\costj$ in the context of \cref{lemma:myerson}. As $\alloc(\bids,\asks)$ for MBT can be implemented by an IC mechanism, by \cref{lemma:myerson}, (a), (b) and (c) hold. Moreover, as ties in utility are broken in favor of a trade, (d) holds.
\end{proofof}

\subsection{Proof of \cref{lemma:allocation_to_payments}}
\label{appsub:proof_of_lemma_allocation_to_payments}

\begin{proofof}{\cref{lemma:allocation_to_payments}}
    For (a), according to \cref{lemma:myerson_with_two_groups} (b),
    \begin{equation*}
        \pay(\beta, \bidsmi, \asks) - \pay(\alpha, \bidsmi, \asks)
        = \beta \cdot \alloc(\beta, \bidsmi, \asks) - \alpha \cdot \alloc(\alpha, \bidsmi, \asks) - \int_{z=\alpha}^{\beta} \alloc(z,\bidsmi,\asks)\d z.
    \end{equation*}
    If $\alloc(\alpha, \bidsmi, \asks) = \alloc(\beta, \bidsmi, \asks)$, by \cref{lemma:myerson_with_two_groups} (a), $\alloc(\bidi, \bidsmi, \asks)$ is constant for $\bidi\in [\alpha,\beta]$. Then according to the equation above, $\pay(\beta, \bidsmi, \asks) - \pay(\alpha, \bidsmi, \asks) = 0$. (a) holds.

    For (b), according to \cref{lemma:myerson_with_two_groups} (c),
    \begin{equation*}
        \rec(\bids, \beta, \asksmj) - \rec(\bids, \alpha ,\asksmj)
        = \beta \cdot \alloc(\bids, \beta, \asksmj) - \alpha \cdot \alloc(\bids, \alpha, \asksmj) + \int_{z=\beta}^{\alpha} \alloc(\bids,z,\asksmj)\d z.
    \end{equation*}
    By similar argument about \cref{lemma:myerson_with_two_groups} (a), we see (b) also holds.
\end{proofof}

\subsection{Proof of \cref{lemma:necessity}}
\label{appsub:proof_of_lemma_necessity}

\begin{proofof}{\cref{lemma:necessity}}
    We will prove both statements by contradiction.

    Suppose (a) does not hold, i.e., $\alloc(\alpha, \bidsmi, \asks) = 0$ but $\alloc(\beta, \bidsmi, \asks) = 1$. According to \cref{lemma:myerson_with_two_groups} (b),
    \begin{equation*}
        \pay(\beta, \bidsmi, \asks) - \pay(\alpha, \bidsmi, \asks)
        = \beta - \int_{z=\alpha}^{\beta} \alloc(z,\bidsmi,\asks)\d z.
    \end{equation*}

    On the other hand, note that as $\alloc(\alpha, \bidsmi, \asks) = 0$ and $\alloc(\beta, \bidsmi, \asks) = 1$, by \cref{lemma:myerson_with_two_groups} (a), $\alloc(\vec 0, \asks) = 0$ and $\alloc(\vec 1, \asks) = 1$, so $\alloc(\alpha, \bidsmi, \asks) = \alloc(\vec 0, \asks)$ and $\alloc(\beta, \bidsmi, \asks) = \alloc(\vec 1, \asks)$. By \cref{coro:allocation_to_payments} (a), $\pay(\beta, \bidsmi, \asks) - \pay(\alpha, \bidsmi, \asks) = \pay(\vec 1, \asks) - \pay(\vec 0, \asks)=\tau_\asks$. Together with the above equation, we know
    \begin{equation*}
        \int_{z=\alpha}^{\beta} \alloc(z,\bidsmi,\asks)\d z = \beta - \tau_\asks.
    \end{equation*}

    By \cref{lemma:myerson_with_two_groups} (a), (d) and $\alloc \in \{0,1\}$, we can conclude that $\alloc(\bidi, \bidsmi, \asks) = \ind{\bidi \geq \tau_\asks}\ (\bidi \in [\alpha,\beta])$. Thus, $\ind{\alpha \geq \tau_\asks} = 0, \ind{\beta \geq \tau_\asks} = 1$. This is a contradiction.

    Similarly, suppose (b) does not hold, i.e., $\alloc(\bids, \alpha, \asksmj) = 1$ and $\alloc(\bids, \beta, \asksmj) = 0$. By \cref{lemma:myerson_with_two_groups} (c),
    \begin{equation*}
        \rec(\bids, \beta, \asksmj) - \rec(\bids, \alpha ,\asksmj)
        = - \alpha + \int_{z=\beta}^{\alpha} \alloc(\bids,z,\asksmj)\d z.
    \end{equation*}

    On the other hand, note that as $\alloc(\bids, \alpha, \asksmj) = 1$ and $\alloc(\bids, \beta, \asksmj) = 0$, by \cref{lemma:myerson_with_two_groups} (a), $\alloc(\bids, \vec 1) = 0$ and $\alloc(\bids, \vec 0) = 1$, so $\alloc(\bids, \beta, \asksmj) = \alloc(\bids, \vec 1)$ and $\alloc(\bids, \alpha, \asksmj) = \alloc(\bids, \vec 0)$. By \cref{coro:allocation_to_payments} (c), $\rec(\bids, \beta, \asksmj) - \rec(\bids, \alpha ,\asksmj) = \rec(\bids, \vec 1) - \rec(\bids, \vec 0) = -\theta_\bids$. Together with the above equation, we know
    \begin{equation*}
        \int_{z=\alpha}^{\beta} \alloc(\bids,z,\asksmj)\d z = \theta_\bids - \alpha.
    \end{equation*}

    By \cref{lemma:myerson_with_two_groups} (a), (d) and $\alloc \in \{0,1\}$, we can conclude that $\alloc(\bids, \askj, \asksmj) = \ind{\askj \leq \theta_\bids}\ (\askj \in [\alpha,\beta])$. Thus, $\ind{\alpha \leq \theta_\bids} = 1, \ind{\beta \leq \theta_\bids} = 0$. This is a contradiction.
\end{proofof}

\subsection{Proof of \cref{lemma:necessity_sbb}}
\label{appsub:proof_of_lemma_necessity_sbb}

\begin{proofof}{\cref{lemma:necessity_sbb}}
    We will prove both statements by contradiction.

    Suppose (a) does not hold, i.e., $\alloc(\alpha, \bidsmi, \asks) = 0$ but $\alloc(\beta, \bidsmi, \asks) = 1$. According to \cref{lemma:myerson_with_two_groups} (b),
    \begin{equation*}
        \pay(\beta, \bidsmi, \asks) - \pay(\alpha, \bidsmi, \asks)
        = \beta - \int_{z=\alpha}^{\beta} \alloc(z,\bidsmi,\asks)\d z.
    \end{equation*}

    On the other hand, note that as $\alloc(\alpha, \bidsmi, \asks) = 0$ and $\alloc(\beta, \bidsmi, \asks) = 1$, by \cref{lemma:myerson_with_two_groups} (a), $\alloc(\vec 0, \vec 1) = 0$ and $\alloc(\vec 1, \vec 0) = 1$, so $\alloc(\alpha, \bidsmi, \asks) = \alloc(\vec 0, \vec 1)$ and $\alloc(\beta, \bidsmi, \asks) = \alloc(\vec 1, \vec 0)$. By \cref{coro:allocation_to_payments} (c), $\pay(\beta, \bidsmi, \asks) - \pay(\alpha, \bidsmi, \asks) = \pay(\vec 1, \vec 0) - \pay(\vec 0, \vec 1)=\tau$. Together with the above equation, we know
    \begin{equation*}
        \int_{z=\alpha}^{\beta} \alloc(z,\bidsmi,\asks)\d z = \beta - \tau.
    \end{equation*}

    By \cref{lemma:myerson_with_two_groups} (a), (d) and $\alloc \in \{0,1\}$, we can conclude that $\alloc(\bidi, \bidsmi, \asks) = \ind{\bidi \geq \tau}\ (\bidi \in [\alpha,\beta])$. Thus, $\ind{\alpha \geq \tau} = 0, \ind{\beta \geq \tau} = 1$. This is a contradiction.

    Similarly, suppose (b) does not hold, i.e., $\alloc(\bids, \alpha, \asksmj) = 1$ and $\alloc(\bids, \beta, \asksmj) = 0$. By \cref{lemma:myerson_with_two_groups} (c),
    \begin{equation*}
        \rec(\bids, \beta, \asksmj) - \rec(\bids, \alpha ,\asksmj)
        = - \alpha + \int_{z=\beta}^{\alpha} \alloc(\bids,z,\asksmj)\d z.
    \end{equation*}

    On the other hand, note that as $\alloc(\bids, \alpha, \asksmj) = 1$ and $\alloc(\bids, \beta, \asksmj) = 0$, by \cref{lemma:myerson_with_two_groups} (a), $\alloc(\vec 0, \vec 1) = 0$ and $\alloc(\vec 1, \vec 0) = 1$, so $\alloc(\bids, \beta, \asksmj) = \alloc(\vec 0, \vec 1)$ and $\alloc(\bids, \alpha, \asksmj) = \alloc(\vec 1, \vec 0)$. By \cref{coro:allocation_to_payments} (c), $\rec(\bids, \beta, \asksmj) - \rec(\bids, \alpha ,\asksmj) = \rec(\vec 0, \vec 1) - \rec(\vec 1, \vec 0) = -\tau$. Together with the above equation, we know
    \begin{equation*}
        \int_{z=\alpha}^{\beta} \alloc(\bids,z,\asksmj)\d z = \tau - \alpha.
    \end{equation*}

    By \cref{lemma:myerson_with_two_groups} (a), (d) and $\alloc \in \{0,1\}$, we can conclude that $\alloc(\bids, \askj, \asksmj) = \ind{\askj \leq \tau}\ (\askj \in [\alpha,\beta])$. Thus, $\ind{\alpha \leq \tau} = 1, \ind{\beta \leq \tau} = 0$. This is a contradiction.
\end{proofof}

\subsection{Proof of \cref{thm:deterministic_characterization}}
\label{appsub:proof_of_thm_deterministic_characterization}

\begin{proofof}{\cref{thm:deterministic_characterization}}
	The necessity $(\implies)$ is implied by \cref{lemma:necessity}.
    
    Therefore, it suffices to prove the sufficiency $(\impliedby)$. For each $\asks\in[0,1]^n$, find $\tau_\asks$ as stated in condition (a), and let $\pay(\bids, \asks) = \tau_\asks \cdot \alloc(\bids, \asks)$. And for each $\bids\in[0,1]^n$, find $\theta_\bids$ as stated in condition (b), and let $\rec(\bids, \asks) = \theta_\bids \cdot \alloc(\bids, \asks)$. We will verify that $(\alloc, \pay, \rec)$ is an IC mechanism.
    
    For buyer $i$ and any $\bidsmi \in [0,1]^{n-1}, \asks \in [0,1]^n$, recall that
    \begin{equation*}
        \utili(\bidi, \bidsmi, \asks) = \vali\cdot\alloc(\bidi, \bidsmi, \asks) - \pay(\bidi, \bidsmi, \asks) = (\vali - \tau_\asks)\cdot \alloc(\bidi, \bidsmi, \asks).
    \end{equation*}

    For fixed $\bidsmi$ and $ \asks$, $\alloc(\bidi, \bidsmi, \asks)$ is a monotone Boolean function of $\ind{\bidi \geq \tau_\asks}$. Verify that $\bidi = \vali$ maximizes the utility when $\alloc(\bidi, \bidsmi, \asks)$ as a function of $\bidi$ is (i) $0$ (ii) $1$ and (iii) $\ind{\bidi \geq \tau_\asks}$.

    For seller $j$ and any $\asksmj \in [0,1]^{n-1}, \bids \in [0,1]^n$, recall that
    \begin{equation*}
        \utilj(\bids, \askj, \asksmj) = \rec(\bids, \askj, \asksmj) - \costj\cdot\alloc(\bids, \askj, \asksmj) = (\theta_\bids - \costj) \cdot \alloc(\bids, \askj, \asks).
    \end{equation*}

    For fixed $\asksmj$ and $\bids$, $\alloc(\bids, \askj, \asksmj)$ is a monotone Boolean function of $\ind{\askj \leq \theta_\bids}$. Verify that $\askj = \costj$ maximizes the utility when $\alloc(\bids, \askj, \asksmj)$ as a function of $\askj$ is (i) $0$ (ii) $1$ and (iii) $\ind{\askj \leq \theta_\bids}$.

    This shows that $(\alloc, \pay, \rec)$ is an IC mechanism, and the sufficiency of  \cref{thm:deterministic_characterization} holds.
\end{proofof}

\subsection{Proof of \cref{thm:deterministic_characterization_sbb}}
\label{appsub:proof_of_thm_deterministic_characterization_sbb}

\begin{proofof}{\cref{thm:deterministic_characterization_sbb}}
	The necessity $(\implies)$ is implied by \cref{lemma:necessity_sbb}.
    
    Therefore, it suffices to prove the sufficiency $(\impliedby)$. Let $\pay(\bids, \asks) = \rec(\bids, \asks) = \tau \alloc(\bids, \asks)$. Clearly the mechanism is SBB. We will verify that $(\alloc, \pay, \rec)$ is an IC mechanism.
    
    For buyer $i$ and any $\bidsmi \in [0,1]^{n-1}, \asks \in [0,1]^n$, recall that
    \begin{equation*}
        \utili(\bidi, \bidsmi, \asks) = \vali\cdot\alloc(\bidi, \bidsmi, \asks) - \pay(\bidi, \bidsmi, \asks) = (\vali - \tau)\cdot \alloc(\bidi, \bidsmi, \asks).
    \end{equation*}

    For fixed $\bidsmi$ and $ \asks$, $\alloc(\bidi, \bidsmi, \asks)$ is a monotone Boolean function of $\ind{\bidi \geq \tau}$. Verify that $\bidi = \vali$ maximizes the utility when $\alloc(\bidi, \bidsmi, \asks)$ as a function of $\bidi$ is (i) $0$ (ii) $1$ and (iii) $\ind{\bidi \geq \tau}$.

    For seller $j$ and any $\asksmj \in [0,1]^{n-1}, \bids \in [0,1]^n$, recall that
    \begin{equation*}
        \utilj(\bids, \askj, \asksmj) = \rec(\bids, \askj, \asksmj) - \costj\cdot\alloc(\bids, \askj, \asksmj) = (\tau - \costj) \cdot \alloc(\bids, \askj, \asks).
    \end{equation*}

    For fixed $\asksmj$ and $\bids$, $\alloc(\bids, \askj, \asksmj)$ is a monotone Boolean function of $\ind{\askj \leq \tau}$. Verify that $\askj = \costj$ maximizes the utility when $\alloc(\bids, \askj, \asksmj)$ as a function of $\askj$ is (i) $0$ (ii) $1$ and (iii) $\ind{\askj \leq \tau}$.

    This shows that $(\alloc, \pay, \rec)$ is an IC mechanism, and the sufficiency of  \cref{thm:deterministic_characterization_sbb} holds.
\end{proofof}

\subsection{Proof of \cref{thm:forced_trade}}
\label{appsub:proof_of_thm_forced_trade}

\begin{proofof}{\cref{thm:forced_trade}}
    Let $\mu_v = \Ex[\val\sim\bdist]{\val},\mu_c = \Ex[\cost\sim\sdist]{\cost}$. Recall that for buyer $i$ and seller $j$,
    \begin{align*}
        \utili(\bids,\asks) &= \alloc(\bids,\asks)\cdot\vali -\pay(\bids,\asks)=v_i-0.5(\mu_v+\mu_c),\\
        \utilj(\bids,\asks) &= \rec(\bids,\asks) - \alloc(\bids,\asks)\cdot\costj = 0.5(\mu_v+\mu_c) - c_j.
    \end{align*}

    The participants' utilities do not depend on their actions, so the mechanism is IC. And as $\pay(\bids,\asks) = \rec(\bids,\asks)$ always holds, we can see that the mechanism is SBB. For IR, compute that
    \begin{align*}
        \sum_{i=1}^{n}\utili(\vals,\costs) &= \sum_{i=1}^{n}\vali - 0.5n(\mu_v+\mu_c),\\
        \sum_{j=1}^{n}\utilj(\vals,\costs) &= 0.5n(\mu_v+\mu_c) - \sum_{j=1}^{n}\costj.\\
    \end{align*}
    Let $\delta = 0.5(\mu_v-\mu_c) > 0$. Using Chernoff bounds, we can see that
    \begin{align}
        \Prx{\sum_{i=1}^{n}\utili(\vals,\costs)\leq 0}
        &\leq \Prx[\vals\sim\bdist^n]{\sum_{i=1}^{n}\vali \leq n(\mu_v - \delta)} \leq e^{-\frac{n\delta^2}{3\mu_v}}, \label{eq:forced_trade1}\\
        \Prx{\sum_{j=1}^{n}\utilj(\vals,\costs)\leq 0}
        &\leq \Prx[\costs\sim\sdist^n]{\sum_{j=1}^{n}\costj \geq n(\mu_c + \delta)} \leq e^{-\frac{n\delta^2}{3\mu_c}}. \label{eq:forced_trade2}
    \end{align}
    Note that $\bdist$ and $\sdist$ are supported on $[0, 1]$, so $\mu_v>\mu_c\geq 0$. According to \eqref{eq:forced_trade1}, $\Prx{\sum_{i=1}^{n}\utili(\vals,\costs)\leq 0} = 1-e^{-\Omega(n)}$, which means IR is satisfied for the buyers with probability $1-e^{-\Omega(n)}$. If $\mu_c>0$, then using \eqref{eq:forced_trade2}, $\Prx{\sum_{j=1}^{n}\utilj(\vals,\costs)\leq 0} = 1-e^{-\Omega(n)}$. Otherwise, $\mu_c=0$, then $\Prx[\cost\sim\sdist]{\cost = 0}=1$, and $\Prx{\sum_{j=1}^{n}\utilj(\vals,\costs)\leq 0} = 0$, which shows that the sellers are IR with probability $1-e^{-\Omega(n)}$.

    Finally, for efficiency, compute that
    \begin{align}
        \FB &= \Ex[\vals \sim \bdist^n, \costs \sim \sdist^n] {
            \InParentheses{\sum_{i} \vali - \sum_{j} \costj} \cdot \ind{\sum_{i} \vali \geq \sum_{j} \costj}
        } \notag\\
        &\leq \Ex[\vals \sim \bdist^n, \costs \sim \sdist^n] {
            \InParentheses{\sum_{i} \vali - \sum_{j} \costj}  + n\cdot \ind{\sum_{i} \vali < \sum_{j} \costj}
        } \notag\\
        &\leq n\InBrackets{\mu_v-\mu_c+\Prx[\vals\sim\bdist^n]{\sum_{i}\vali \leq n(\mu_v - \delta)} + \Prx[\costs\sim\sdist^n]{\sum_{j}\costj \geq n(\mu_c + \delta)}} \notag\\
        &\leq n\InBrackets{\mu_v-\mu_c+e^{-\frac{n\delta^2}{3\mu_v}} + e^{-\frac{n\delta^2}{3\mu_c}}} \label{eq:forced_trade3}
    \end{align}
    Here, the final inequality uses Chernoff bounds. And $\ALG = \Ex[\vals \sim \bdist^n, \costs \sim \sdist^n] {
        \InParentheses{\sum_{i} \vali - \sum_{j} \costj} \cdot 1
    } = n(\mu_v - \mu_c)$. If $\mu_c=0$, then $\Prx[\cost\sim\sdist]{\cost = 0}=1$, and $\ALG = \FB = n(\mu_v-\mu_c)$ must hold. Otherwise, according to \eqref{eq:forced_trade3}, we can see that $\frac{\ALG}{\FB}=1-e^{-\Omega(n)}$. This concludes the proof.
\end{proofof}

\subsection{Proof of \cref{thm:deterministic_hardness}}
\label{appsub:proof_of_thm_deterministic_hardness}

\begin{proofof}{\cref{thm:deterministic_hardness}}
    We will prove this theorem by showing that under the characterization of IC mechanisms in \cref{thm:deterministic_characterization}, no IC mechanisms can be efficient in the case where $\bdist$ is a uniform distribution on $[0, 1]$ and $\sdist$ is always $\frac{1}{2}$. Note that $\Ex[\val\sim\bdist]{\val}=\Ex[\cost\sim\sdist]{\cost}=\frac{1}{2}$ in this case.

    We first compute that $\FB = \Ex[\vals \sim \U{[0, 1]^n}] {
        \InParentheses{\sum_{i} \vali - \frac{1}{2}n} \cdot \ind{\sum_{i} \vali \geq \frac{1}{2}n}
    }$. Let $\mu_n = \frac{1}{n} \sum_{i} \vali$, then $\FB = n \cdot \Ex[\vals \sim \U{[0, 1]^n}] {(\mu_n - \frac{1}{2}) \cdot \ind{\mu_n \geq \frac{1}{2}}}$. Note that for $i\in [n]$, $\vali$ are i.i.d$.$ random variables with mean $\mu = \frac{1}{2}$ and variance $\sigma^2 = \frac{1}{12}$. Using the central limit theorem, $\sqrt{n}(\mu_n - \frac{1}{2})$ converges in distribution to a normal distribution $\N(0,\frac{1}{12})$. Thus,
    \begin{align}
        \FB = (1 + o(1))\sqrt{n}\cdot \Ex[x\sim \N(0,\frac{1}{12})]{x \cdot \ind{x\geq 0}} = (1 + o(1))\sqrt{\frac{n}{24\pi}}. \label{eq:deterministic_hardness1}
    \end{align}

    On the other hand, consider an IC mechanism $(\alloc, \pay, \rec)$. As $\costs$ is constant and the mechanism is IC, $\asks$ is also constant, and thus $\alloc$ only depends on $\bids$. Then according to \cref{thm:deterministic_characterization}, there exist $\tau \in [0,1]$ and a monotone Boolean function $f:\{0,1\}^{n}\to\{0,1\}$, such that $\alloc(\bids) = f(\ind{\bidi[1]\geq\tau}, \cdots, \ind{\bidi[n]\geq\tau})$. Therefore,
    \begin{equation*}
        \ALG = \Ex[\vals \sim {\U[0, 1]^n}] {
            \InParentheses{\sum_{i} \vali - \frac{1}{2}n} \cdot f\InParentheses{\ind{\vali[1]\geq\tau}, \cdots, \ind{\vali[n]\geq\tau}}
        }.
    \end{equation*}

    Let $e_i = \ind{\vali \geq \tau}$ and $s = \sum_{i=1}^{n}e_i$. Compute that $\Ex{ \InParentheses{\sum_{i} \vali - \frac{1}{2}n} \mid \mathbf e} = \frac{1}{2}(s - (1 - \tau) n)$. Therefore, to maximize $\ALG$, $f(\mathbf e) = \ind{s \geq (1 - \tau) n}$. Let $m = (1 - \tau) n$,\footnote{For simplicity of presentation, assume $\tau n$ is an integer to avoid rounding issues.} we can compute that
    \begin{align}
        \ALG &\leq \sum_{i=m}^{n}\Prx{s = i}\cdot \frac{1}{2}(i - (1 - \tau) n)\notag \\
        & = \sum_{i=m}^{n} \tau^{n - i} (1 - \tau)^i \binom{n}{i} \cdot \frac{1}{2}(i - (1 - \tau) n)\notag \\
        & = \frac{n}{2} \sum_{i=m}^{n} \InParentheses{\tau^{n - i} (1 - \tau)^i \binom{n - 1}{i - 1} - \tau^{n - i} (1 - \tau)^{i + 1} \binom{n}{i}}\notag \\
        & = \frac{n}{2}\cdot \tau^{n - m + 1}(1-\tau)^{m}\binom{n - 1}{m - 1} \label{eq:deterministic_hardness2}\\
        & = \frac{n}{2}\cdot \tau^{n - m + 1}(1-\tau)^{m + 1}\binom{n}{m} \label{eq:deterministic_hardness3}
    \end{align}

    Here, \eqref{eq:deterministic_hardness2} depends on the fact that $\binom{n}{i} = \binom{n - 1}{i - 1} + \binom{n - 1}{i}$ and \eqref{eq:deterministic_hardness3} is because $m = (1 - \tau)n$. 

    Applying Stirling's approximation to \eqref{eq:deterministic_hardness3}, we see that
    \begin{equation}
        \ALG \leq (1 + o(1))\sqrt{\frac{n}{8\pi}\tau(1-\tau)}\leq (1 + o(1))\sqrt{\frac{n}{32\pi}}.\label{eq:deterministic_hardness4}
    \end{equation}
    Combining \eqref{eq:deterministic_hardness1} and \eqref{eq:deterministic_hardness4}, we can see that $\frac{\ALG}{\FB}\leq \frac{\sqrt{3}}{2} + o(1)$, which concludes the proof.
\end{proofof}

\section{Missing Proofs in Section \ref{sec:randomized_mechanisms}}
\label{app:randomized_mechanisms}

\subsection{Proof of \cref{lemma:calculus}}
\label{appsub:proof_of_lemma_calculus}

\begin{proofof}{\cref{lemma:calculus}}
    (i) $\implies$ (ii): Directly compute that $\frac{\partial f}{\partial x_i}(\mathbf x) = f'_i(x_i)$ and $\frac{\partial^2 f}{\partial x_i\partial x_j}(\mathbf x) = 0$.

    (ii) $\implies$ (i): Using (ii), we can first see that for any $i,j\in[n], i\ne j$, $\frac{\partial f}{\partial x_i}(x_j,\mathbf x_{-j})=\frac{\partial f}{\partial x_i}(x_j',\mathbf x_{-j})$. Applying this multiple times, we further know that for any $i\in [n]$, $\frac{\partial f}{\partial x_i}(x_i,\mathbf x_{-i})=\frac{\partial f}{\partial x_i}(x_i,\mathbf x_{-i}')$. Then, for each $i\in [n]$, take an arbitrary $x_{-i}'$, let $f_i(x_i) = \int_{i=0}^{x_i}\frac{\partial f}{\partial x_i}(x_i,\mathbf x_{-i})$ and let $g(\mathbf x) = f(\vec 0) + \sum_{i=1}^{n}f_i(x_i)$. We can see that $g(\vec 0) = f(\vec 0)$ and for each $i\in[n]$, $\frac{\partial g}{\partial x_i}(\mathbf x) = \frac{\partial f}{\partial x_i}(\mathbf x)$. This shows that $f(\mathbf x) = g(\mathbf x) = f(\vec 0) + \sum_{i=1}^{n}f_i(x_i)$, which concludes the proof.
\end{proofof}

\subsection{Proof of \cref{lemma:independent}}
\label{appsub:proof_of_lemma_independent}

\begin{proofof}{\cref{lemma:independent}}
    For (a), according to \cref{lemma:myerson_with_two_groups} (b),
    \begin{equation}
        \pay(\beta_i, \bidsmi, \asks) - \pay(\alpha, \bidsmi, \asks)
        = \beta_i \cdot \alloc(\beta_i, \bidsmi, \asks) - \alpha_i \cdot \alloc(\alpha_i, \bidsmi, \asks) - \int_{z=\alpha_i}^{\beta_i} \alloc(z,\bidsmi,\asks)\d z. \label{eq:independent1}
    \end{equation}
    Plugg \eqref{eq:independent1} into the following identical equation:
    \begin{align*}
        &[
            p(\beta_{i1},\beta_{i2},\bidsmi[i1,i2],\asks)
            -p(\beta_{i1},\alpha_{i2},\bidsmi[i1,i2],\asks)
        ]
        +
        [
            p(\beta_{i1},\alpha_{i2},\bidsmi[i1,i2],\asks)
            -p(\alpha_{i1},\alpha_{i2},\bidsmi[i1,i2],\asks)
        ]\\
        =\ 
        &[
            p(\beta_{i1},\beta_{i2},\bidsmi[i1,i2],\asks)
            -p(\alpha_{i1},\beta_{i2},\bidsmi[i1,i2],\asks)
        ]
        +
        [
            p(\alpha_{i1},\beta_{i2},\bidsmi[i1,i2],\asks)
            -p(\alpha_{i1},\alpha_{i2},\bidsmi[i1,i2],\asks)
        ].
    \end{align*}
    We will get \eqref{eq:independent2}. We omit $\bidsmi[i1,i2],\asks$ in \eqref{eq:independent2} for readability.
    \begin{align}
        &\beta_{i2} \cdot \alloc(\beta_{i1},\beta_{i2})-\alpha_{i2} \cdot \alloc(\beta_{i1},\alpha_{i2})-\int_{\alpha_{i2}}^{\beta_{i2}}\alloc(\beta_{i1},z)\d z \notag\\
        +\ &\beta_{i1} \cdot \alloc(\beta_{i1},\alpha_{i2})-\alpha_{i1} \cdot \alloc(\alpha_{i1},\alpha_{i2})-\int_{\alpha_{i1}}^{\beta_{i1}}\alloc(z,\alpha_{i2})\d z \notag\\
        =\ &\beta_{i1} \cdot \alloc(\beta_{i1},\beta_{i2})-\alpha_{i1} \cdot \alloc(\alpha_{i1},\beta_{i2})-\int_{\alpha_{i1}}^{\beta_{i1}}\alloc(z,\beta_{i2})\d z \notag\\
        +\ &\beta_{i2} \cdot \alloc(\alpha_{i1},\beta_{i2})-\alpha_{i2} \cdot \alloc(\alpha_{i1},\alpha_{i2})-\int_{\alpha_{i2}}^{\beta_{i2}}\alloc(\alpha_{i1},z)\d z. \label{eq:independent2}
    \end{align}
    In \eqref{eq:independent2}, take the derivative of $\beta_{i1}$, and then take the derivative of $\beta_{i2}$. We know that
    \begin{equation*}
        \beta_{i2}\cdot \frac{\partial^2 \alloc}{\partial \bidi[i1] \partial \bidi[i2]}(\beta_{i1}, \beta_{i2}, \bidsmi[i1,i2],\asks)=\beta_{i1}\cdot \frac{\partial^2 \alloc}{\partial \bidi[i1] \partial \bidi[i2]}(\beta_{i1}, \beta_{i2}, \bidsmi[i1,i2],\asks).
    \end{equation*}

    This shows that when $\beta_1\ne \beta_2$, $\frac{\partial^2 \alloc}{\partial \bidi[i1] \partial \bidi[i2]}(\beta_{i1}, \beta_{i2}, \bidsmi[i1,i2],\asks) = 0$. Using the continuity of $\frac{\partial^2 \alloc}{\partial \bidi[i1]\partial \bidi[i2]}$, we further know that $\frac{\partial^2 \alloc}{\partial \bidi[i1]\partial \bidi[i2]}(\bids,\asks) = 0$, i.e., (a) holds.

    For (b), according to \cref{lemma:myerson_with_two_groups} (c),
    \begin{equation}
        \rec(\bids, \beta_j, \asksmj) - \rec(\bids, \alpha_j ,\asksmj)
        = \beta_j \cdot \alloc(\bids, \beta_j, \asksmj) - \alpha_j \cdot \alloc(\bids, \alpha_j, \asksmj) + \int_{z=\beta_j}^{\alpha_j} \alloc(\bids,z,\asksmj)\d z. \label{eq:independent3}
    \end{equation}
    With an argument similar to (a), we can prove (b) also holds. 
\end{proofof}

\subsection{Proof of \cref{thm:randomized_characterization}}
\label{appsub:proof_of_thm_randomized_characterization}

\begin{proofof}{\cref{thm:randomized_characterization}}
    The necessity $(\implies)$ is implied by \cref{lemma:calculus,lemma:independent}, \cref{lemma:myerson_with_two_groups} (a).
    
    Therefore, it suffices to prove the sufficiency $(\impliedby)$. For each $\asks\in[0,1]^n$, find functions $f_{\asks,i}$ as stated in condition (a), and let $\pay(\bids, \asks) = \sum_{i=1}^{n}\InParentheses{\bidi\cdot f_{\asks,i}(\bidi) - \int_{z=0}^{\bidi}f_{\asks,i}(z)\d z}$. And for each $\bids\in[0,1]^n$, find $g_{\bids,j}$ as stated in condition (b), and let $\rec(\bids, \asks) = \sum_{j=1}^{n}\InParentheses{\askj\cdot g_{\bids,j}(\askj)+\int_{z=\askj}^{1}g_{\bids,j}(z)\d z}$.
    
    We will verify that $(\alloc, \pay, \rec)$ is an IC mechanism.

    For buyer $i$, fix $\bidsmi \in [0,1]^{n-1},\asks \in [0,1]^n$, recall that
    \begin{equation*}
        \utili(\bidi) = \vali\cdot\alloc(\bidi) - \pay(\bidi) = \vali\sum_{i=1}^{n}f_{\asks,i}(\bidi) - \sum_{i=1}^{n}\InParentheses{\bidi\cdot f_{\asks,i}(\bidi) - \int_{z=0}^{\bidi}f_{\asks,i}(z)\d z}.
    \end{equation*}

    Then $\utili'(\bidi) = (\vali - \bidi) f_{\asks,i}'(\bidi)$. As $f_{\asks,i}'\geq 0$, bidding $\bidi = \vali$ maximizes buyer $i$'s utility.

    For seller $j$, fix $\bids \in [0,1]^{n},\asksmj \in [0,1]^{n-1}$, recall that
    \begin{equation*}
        \utilj(\askj) = \rec(\askj) - \costj\cdot\alloc(\costj) = \sum_{j=1}^{n}\InParentheses{\askj\cdot g_{\bids,j}(\askj) - \int_{z=\askj}^{1}g_{\bids,j}(z)\d z} - \costj\sum_{j=1}^{n}g_{\bids,j}(\costj).
    \end{equation*}

    Then $\utilj'(\askj) = (\askj - \costj) g_{\bids,j}'(\askj)$. As $g_{\bids,j}'\leq 0$, asking $\askj = \costj$ maximizes seller $j$'s utility.

    This concludes the proof of \cref{thm:randomized_characterization}.
\end{proofof}

\subsection{Proof of \cref{thm:randomized_hardness}}
\label{appsub:proof_of_thm_randomized_hardness}

\begin{proofof}{\cref{thm:randomized_hardness}}
    We will prove this theorem by showing that under the characterization of twice continuously differentiable randomized IC mechanisms in \cref{thm:randomized_characterization}, no twice continuously differentiable randomized IC mechanisms can be a constant approximation of $\FB$ on the case where $\bdist$ is a uniform distribution on $[0, 1]$ and $\sdist$ is always $\frac{1}{2}$.

    In the proof of \cref{thm:deterministic_hardness}, we have already shown in \eqref{eq:deterministic_hardness1} that
    \begin{equation*}
        \FB = (1 + o(1))\sqrt{\frac{n}{24\pi}}.
    \end{equation*}
    
    Consider an IC mechanism $(\alloc, \pay, \rec)$. As $\costs$ is constant and the mechanism is IC, $\asks$ is also constant, and thus $\alloc$ only depends on $\bids$. Then according to \cref{thm:randomized_characterization}, there are non-decreasing differentiable functions $f_i:[0,1]\to\R,\forall i \in [n]$ such that $\alloc(\bids,\asks) = f_1(\bidi[1]) + f_2(\bidi[2]) + \cdots + f_n(\bidi[n])$. Thus,
    \begin{equation*}
        \ALG = \Ex[\vals \sim {\U[0, 1]^n}] {
            \sum_{i1}\InParentheses{\vali[i1] - \frac{1}{2}} \cdot \sum_{i2} f_{i2}(\vali[i2])
        }.
    \end{equation*}
    Note that for $i_1\ne i_2$, $\Ex[\vals \sim {\U[0, 1]^n}] {
        \InParentheses{\vali[i1] - \frac{1}{2}} \cdot f_{i2}(\vali[i2])
    } = 0$; we further know that
    \begin{align*}
        \ALG &= \Ex[\vals \sim {\U[0, 1]^n}] {
            \sum_{i}\InParentheses{\vali - \frac{1}{2}} \cdot f_{i}(\vali)
        } = \sum_{i} \Ex[\vali \sim {\U[0, 1]}]{
            \InParentheses{\vali - \frac{1}{2}} \cdot f_{i}(\vali)
        } \\
        &\leq \sum_{i} \Ex[\vali \sim {\U[0, 1]}]{
            \InParentheses{\vali - \frac{1}{2}} \cdot \InParentheses{f_{i}(1)\cdot\ind{\vali \geq \frac{1}{2}} + f_{i}(0)\cdot\ind{\vali < \frac{1}{2}}}
        }\\
        &= \sum_{i} \InParentheses{f_i(1)\cdot \frac{1}{8} - f_i(0)\cdot \frac{1}{8}}\\
        &= \frac{1}{8}\InParentheses{\alloc(\vec 1) - \alloc(\vec 0)} \leq \frac{1}{8}.
    \end{align*}

    This means that $\frac{\ALG}{\FB}\leq (1+o(1))\sqrt{\frac{3\pi}{8n}}$, i.e., $\ALG$ is not a constant approximation of $\FB$.
\end{proofof}

\end{document}